\documentclass[12pt]{article}
\usepackage{graphicx}
\usepackage{cite}

\graphicspath{{./}{Figs/}}

\begin{document}

\title{Identifying a would-be terrorist:  \\An ineradicable error in the data processing\\?}

\author{Serge Galam\\ CEVIPOF - Centre for Political Research,\\
SciencesPo and CNRS, \\ 1, Place Saint-Thomas d'Aquin, 75007 Paris, France\\
serge.galam@sciencespo.fr}
\date{Chaos, Solitons \& Fractals,168, 113119 (2023)}

\maketitle

\begin{abstract}

Processing fragments of data collected on a monitored person to find out whether this person is a would-be terrorist (WT) is very challenging. Moreover, the process has proven to be deceptive, with repeated dramatic failures. To address the issue I suggest a mirror simple model to mimic the process at stake. The model considers a collection of ground items which are labelled either Terrorist Connected (TC) or Terrorist Free (TF). To extract the signal from the ground data items I implement an iterated coarse-grained scheme, which yields a giant unique item with a label TC or TF. The results obtained validate the processing scheme with correct outcomes for the full range of proportions of TC items, beside in a specific sub-range. There, a systematic wrong labelling of the giant item is obtained at the benefit of WT, who are wrongly labeled not would-be terrorist (NWT). This flaw proves to be irremovable because it is anchored within the processing itself in connexion with the treatment of uncertain aggregates, which inevitable appear. The ``natural" allocation of uncertain aggregates to the TF label, in tune with the ethical application of the presumption of innocence in force in democracies, confines the wrong diagnosis systematically to the benefit of some WT, who are labelled NWT. It happens that applying the presumption of guilt, instead of that of innocence, to label the indeterminate aggregates, shifts the systematic errors to another sub-range of proportions of TC items with then zero errors in identifying a WT.  But instead, some NWT are wrongly labeled WT. Despite of being quite far from reality, my results suggest that, just in case, intelligence agencies should investigate the overall effect of seemingly inconsequential "natural" labelling that, applied locally when data are uncertain, are without any immediate detectable impact on the data processing. If the bias proves to be true, making public the choice of presumption of innocence, could avoid the misunderstanding of people when dramatic mistakes are repeated.

\end{abstract}


\section{Introduction}

Identifying a potential terrorist by collecting data on a specific individual is a critical counterterrorism tool to which significant effort is devoted \cite{counter}. And yet, a series of terrorist attacks have not been thwarted despite the covert surveillance of the terrorists involved prior to the attacks as dramatically illustrated with the 2015 terrorist attack in Paris and the 2016 attack in Brussels. Indeed, prior to the attacks almost all of the terrorists involved had been under surveillance only to be eventually declared not to pose a security threat \cite{paris, brus1, brus2}.

Most explanations for these failures have pointed to the need for additional resources and personnel. According to intelligence agencies and security experts, monitoring one individual 24 hours a day means mobilizing a large workforce, making it de facto impossible to monitor hundreds of suspects simultaneously for several weeks in a row. \cite{24}.

More precisely, the collection of information on a given suspect is done through two simultaneous channels \cite{m}. The first channel is directed surveillance that tracks the suspect's conversations,  movements, and other activities. It is operated by specialized people who gather information by direct observation. The second channel, called intrusive surveillance, uses technical devices that secretly collect information about the suspect, including its activities on social networks.

Data collected in these two ways are multiple, scattered, incomplete and can also be contradictory and inconsistent, with different scales and contents. Accordingly, standard machine learning classification algorithms are widely used with high computational power. However, it is an open question whether there are "inconsistent biases" in these standard machine learning algorithms. In addition, the coding of information collected directly can also be subject to caution. To date, no 100\% reliable procedure is available.

Therefore, even when processing a large amount of data using powerful computers, the signal obtained may be false. In particular, since this signal does not exist as such. It must be inferred  from the data processing \cite{public, ba, po}.

On this basis, I raise the possibility of the existence of an intrinsic flaw embedded in the data processing itself, which would make diagnostic errors inevitable, regardless of the adequacy of resources and personnel.

To explore this avenue, given the sensitivity and secrecy of the subject matter, I build an ideal stylized model to mimic the data processing which is at stake. The model consists of an assembly of items labeled respectively "Terrorism Connected" (TC) or "Terrorism Free" (TF), which can be processed simultaneously at maximum by groups of $r$ items. A coarse-grained bottom-up iterative scheme is then implemented with at each step, groups of $r$ items being transformed into single larger items.  Successive enlarged items are labeled TC or TF based on the respective local majorities of their constituent items. 

In case of local ties for even values of $r$, the resulting item is ``naturally" tagged as TF in tune with the ethical values of democratic societies enforcing the presumption of innocence.  In addition, the allocation of  uncertain aggregates at the benefit of TF is perceived as inconsequential among the multitude of repeated labelings that result from local majorities. The process ends with a single giant item labeled either TC (the monitored person is WT) or TF (the  monitored person is NWT). 

The calculations unveil an ineradicable bias with a systematic error in labelling NWT agents who are WT within a well defined range of proportions of ground TC items. Conversely, the model shows that applying the presumption of guilt to tag uncertain aggregates ensures a zero error in the identification of WT. However, the counterpart is a wrong labelling of NWT as WT, which is contrary to the ethical principles of the state of law of democratic societies.

This work subscribes to the current effort to model radicalism \cite{radi1, rad1, rad2, radi2, nuno2}, which is part of the field of sociophysics \cite{brazil, frank, book, fortu, bikas, rum, andre, nuno1, kasia, bello, bolek, cheon, celia, fasa, kk, zaf, vaz, glob, bis}.

The rest of the paper is organized as follows: Section 2 presents the stylized mirror model and the related coarse-grained aggregation scheme is developed in Section 3. An analytical treatment is implemented in Section 4. The Equations are first solved for $r=4$ and then extended to include the combination of different sizes of aggregates.  An ineradicable wrong labelling is found in a specific range of the proportion of TC ground items. Similarity and differences of the present model with previous Galam models are discussed in Section 5 while Section 6 is devoted to a large scale simulation. The cases of wrong conclusions about Weapons of Mass Destruction (WMD) in both Libya and Iraq are briefly revisited in light of the model findings in Section 7. Concluding remarks are contained in last Section.


\section{A stylized mirror model}

I assume that data collected during the monitoring of a suspect agent are represented by two-state items. These items denoted ground items are labeled Terrorist Free (TF) when they could be performed by any person during regular daily activities without any possible connection to terrorism. In contrast, some items are compatible with activities of a Would be Terrorist (WT). Those items are labeled Terrorist Connected (TC).

In addition, I hypothesize that more than fifty percent of TC ground items report a WT. Otherwise, the person is NWT. While the choice of a fifty percent threshold is reasonable, it could be modified without loss of generality.  

Moreover, to account for the fundamental difficulty of processing the ground data that are highly heterogeneous and diverse in nature, I consider that it is not possible to process all the items collected at once in order to determine the respective shares of the TF and TC items. Furthermore, I limit their synthesis by groups of $r$ at most. This restriction makes impossible to average all the items at once to reach a direct evaluation of the proportion of TC items. 

Instead, an iterated bottom-up coarse-grained scheme must be implemented \cite{coarse}. At each step, $r$ items are turned into a single larger item.  The enlarged items are labeled TC or TF according to the respective local majorities of the items that make them up.

In case of equality between TC and TF items in an aggregate with even values of $r$, its classification is uncertain. In this case, the related aggregate is  "naturally" labeled TF without justification, this choice being in line with the ethical values of democracies that enforce the presumption of innocence. \cite{presum}.

However, to gain in generality I tag TC the uncertain aggregate with a probability $k$ and TF with probability $(1-k)$ \cite{hetero}. The value $k=0$ corresponds to an uncertainty breaking at the full benefit of the presumption of innocence. In contrast, $k=1$ corresponds to an uncertainty breaking at the full benefit of the presumption of guilt.

The process is iterated from ground items to level-1 items, which in turn are aggregated and processed into level-2 items and so on until reaching at level $n$ a single giant item, which encompasses the totality of the ground items. If the giant item is labelled TC, the person is WT while it is NWT for a label TF. The value of n is a function of both, the initial proportions of TF and TC of ground items and the size $r$ used to aggregate and synthesize the items. 

At this stage, it is worth to highlight a major difference between the tie breaking effect at a trial and here. In a trial, the doubt may occur at the end of the process when the jury must decide on the guilt or innocence of the accused person. This tie breaking is decisive and final after the completion of the entire trial. People are aware of it and its consequences.

On the other hand, in the case of data processing, tie breaking occurs here and there at different levels without immediate detectable effect. Those cases are thus not given much attention. 


\section{The coarse-grained aggregation scheme}

\begin{figure}[t]
\centering
\includegraphics[width=.80\textwidth]{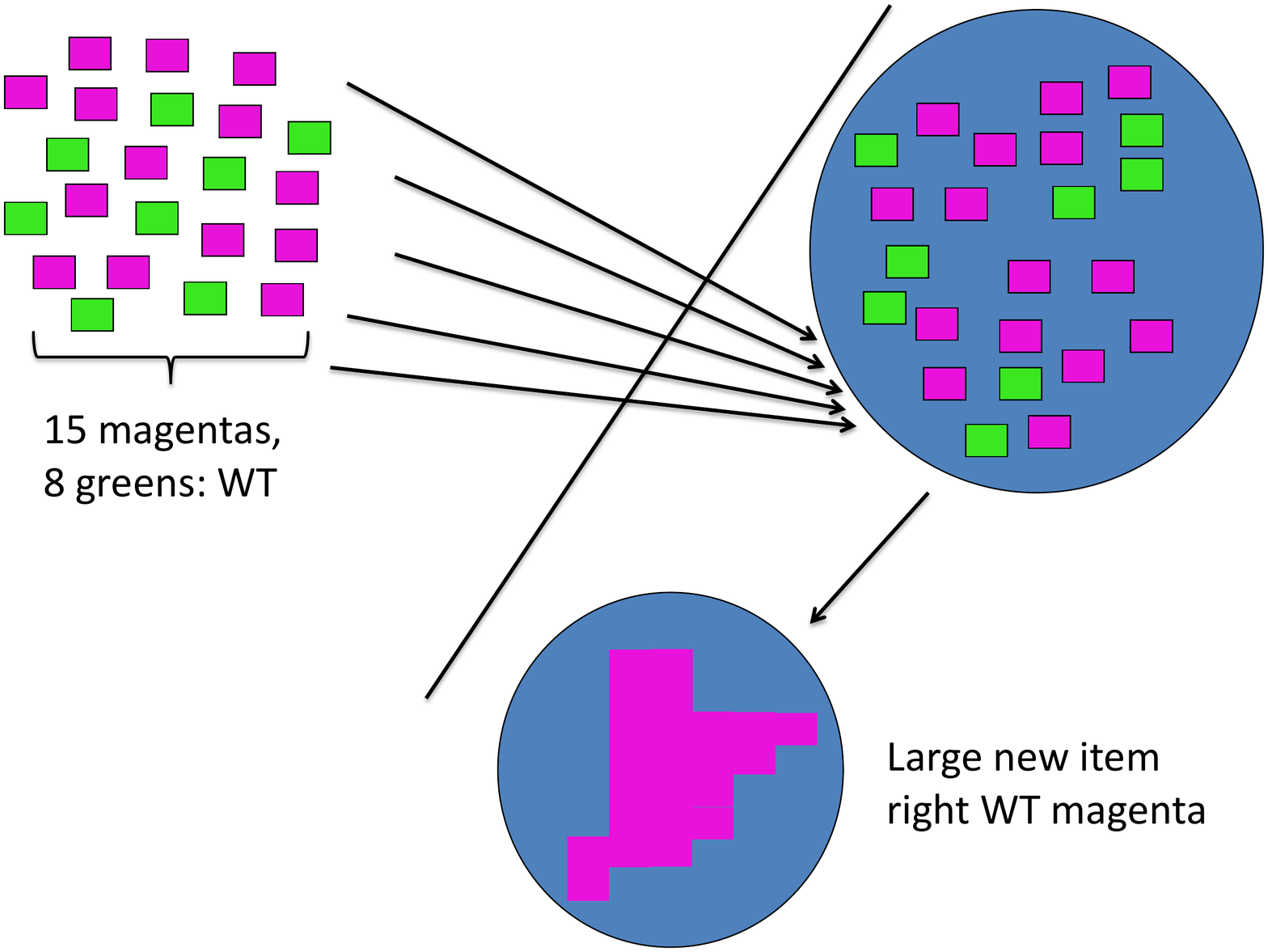}
\caption{A collection of 23 ground items with 12 TF (green) and 11 TC (magenta). One large and unique coarse-grained reveals the right TF label. However, this one-step process is not feasible here because a maximum of $7$ items can be processed together.}
\label{puzzle-all1}
\end{figure}

\begin{figure}[t]
\centering
\includegraphics[width=.80\textwidth]{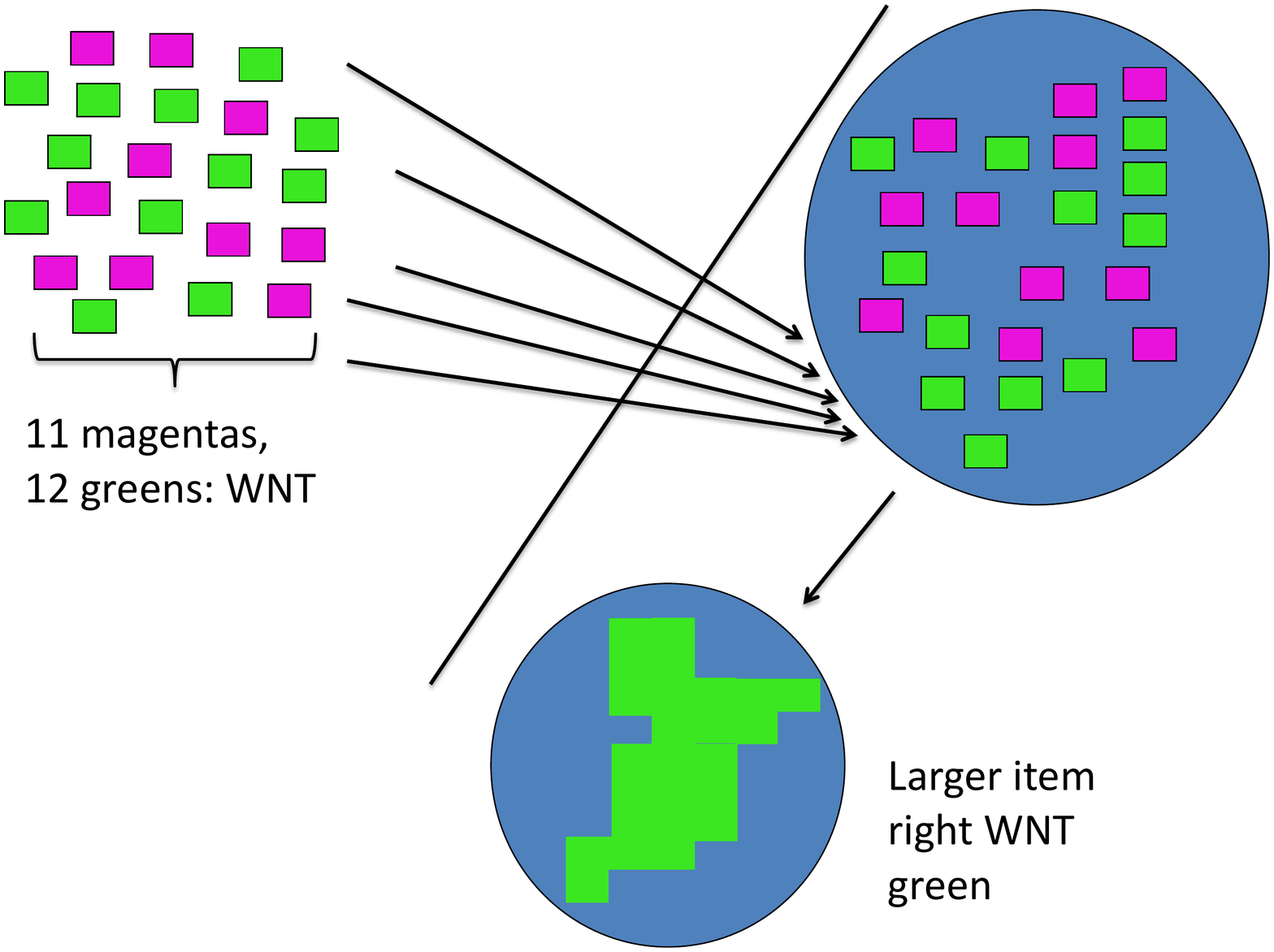}
\caption{A collection of 23 items with 8 TF (green) and 15 TC (magenta). One large and unique coarse-grained reveals the right WT label. However, this one-step process is not feasible here because a maximum of $7$ items can be processed together.}
\label{puzzle-all2}
\end{figure}

To illustrate the coarse-grained scheme I first show a sample of 23 items with 11 TC (magenta) and 12 TF (green) with $r=7$. The majority of TF items reports a NWT as shown in Figure (\ref{puzzle-all1}). In contrast, Figure (\ref{puzzle-all2}) shows another sample of 23 items but now with 15 TC (magenta) and 8 TF (green), which reports a WT.

\begin{figure}
\includegraphics[width=1\textwidth]{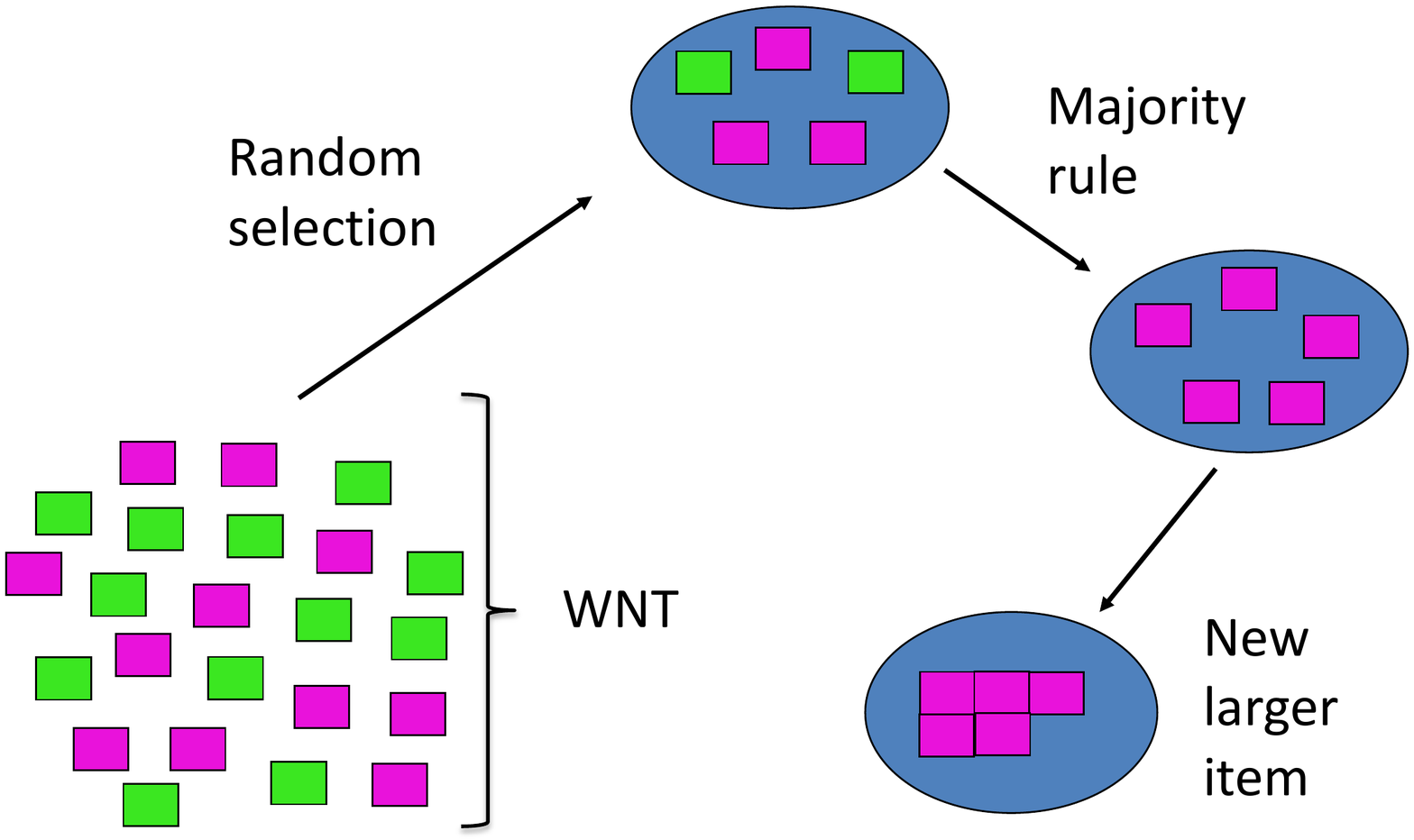}
\caption{A collection of 23 items with 11 TC (magenta) and 12 TF (green) meaning the related person is NWT. The selected group of five items has 3 TC (magenta) and 2 TF (green).  The coarse-grained level-1 item is then tagged TC from majority rule, which is a wrong result.}
\label{impair}
\end{figure}

\begin{figure}
\includegraphics[width=1\textwidth]{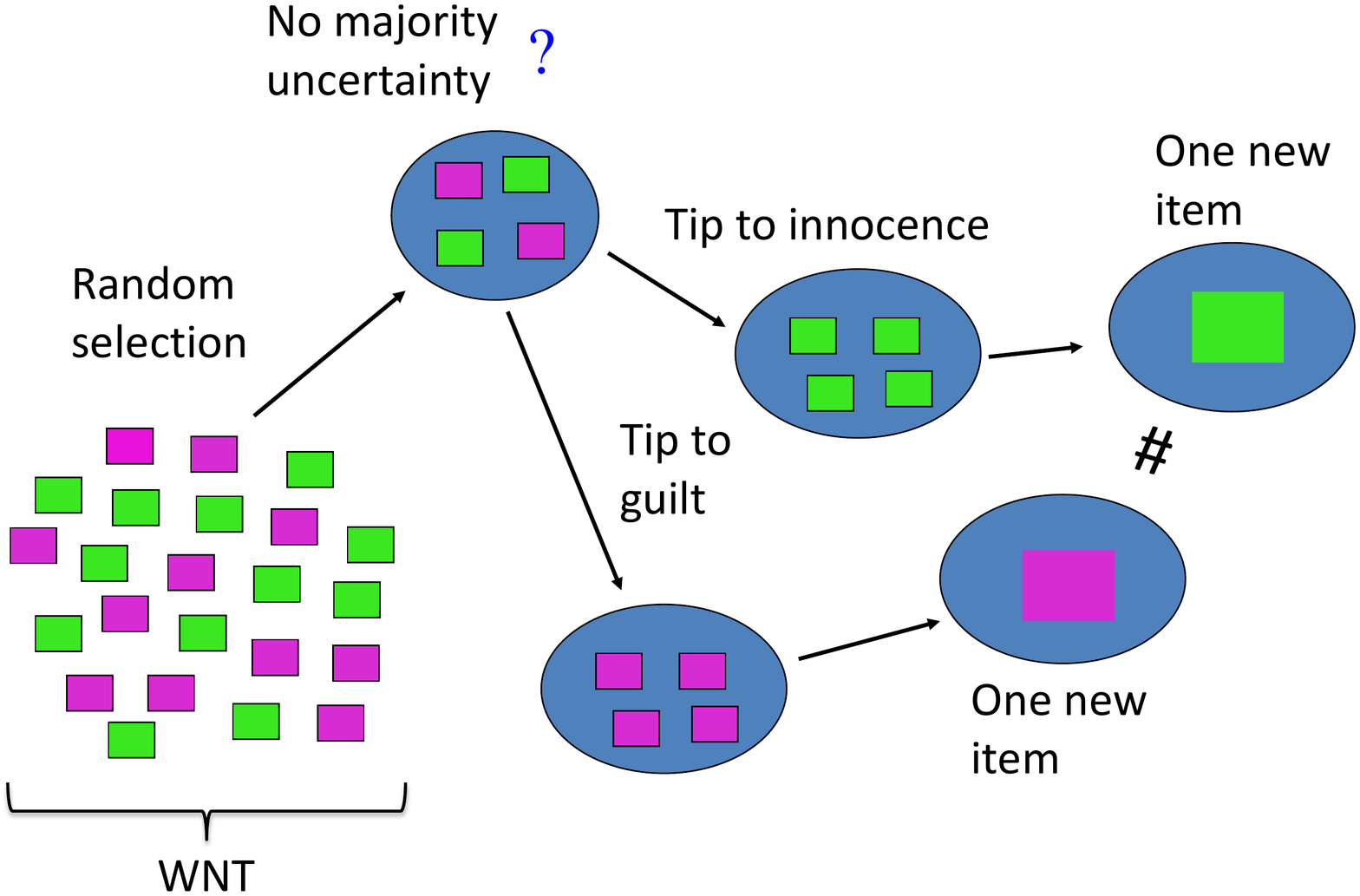}
\caption{A collection of 23 items with 11 TC (magenta) and 12 TF (green) meaning the related person is NWT. A selection of 4 items with 2 TC (magenta) and 2 TF (green) yields one uncertain synthesis (no majority). The related coarse-grained level-1 item is tagged TF (green) for a presumption of innocence ($k=0$ ) or TC (magenta) for presumption of guilt ($k=1$). Here the presumption of innocence yields the right result.}
\label{pair}
\end{figure}

However as stated before, to account for the diversity and heterogeneity of collected items I am assuming that items cannot be coarse-grained at once into one large and unique item. Only small groups of items can be treated together as exhibited in Figures (\ref{impair}, \ref{pair}) with one group of respectively five and four ground items. First case yields a wrong WT label (magenta). Second case leads a right NWT (green) with the uncertain aggregate labelled TF. If tagged TC, the conclusion is WT (magenta), which is wrong.

\begin{figure}
\includegraphics[width=.60\textwidth]{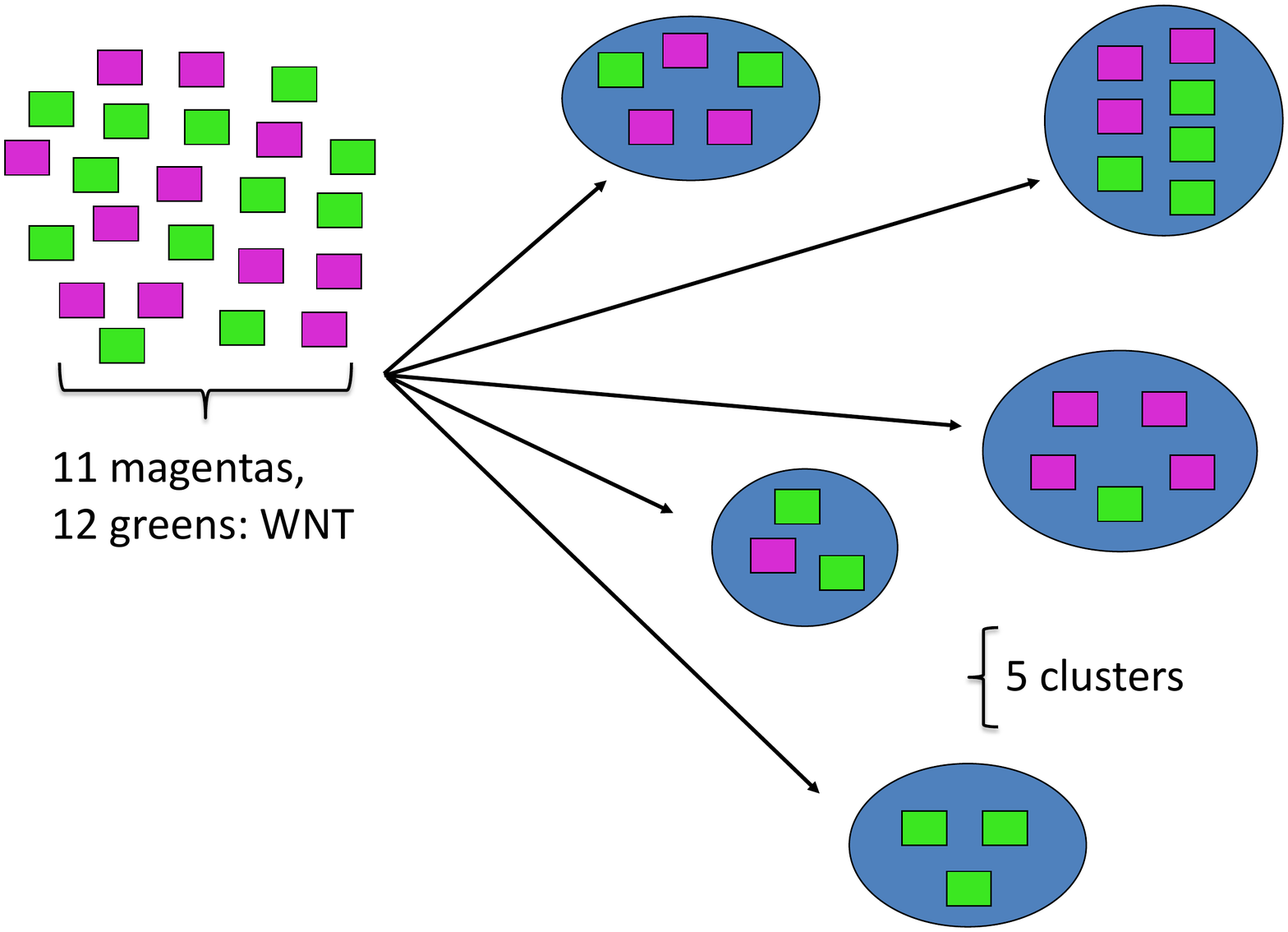}\quad
\includegraphics[width=.50\textwidth]{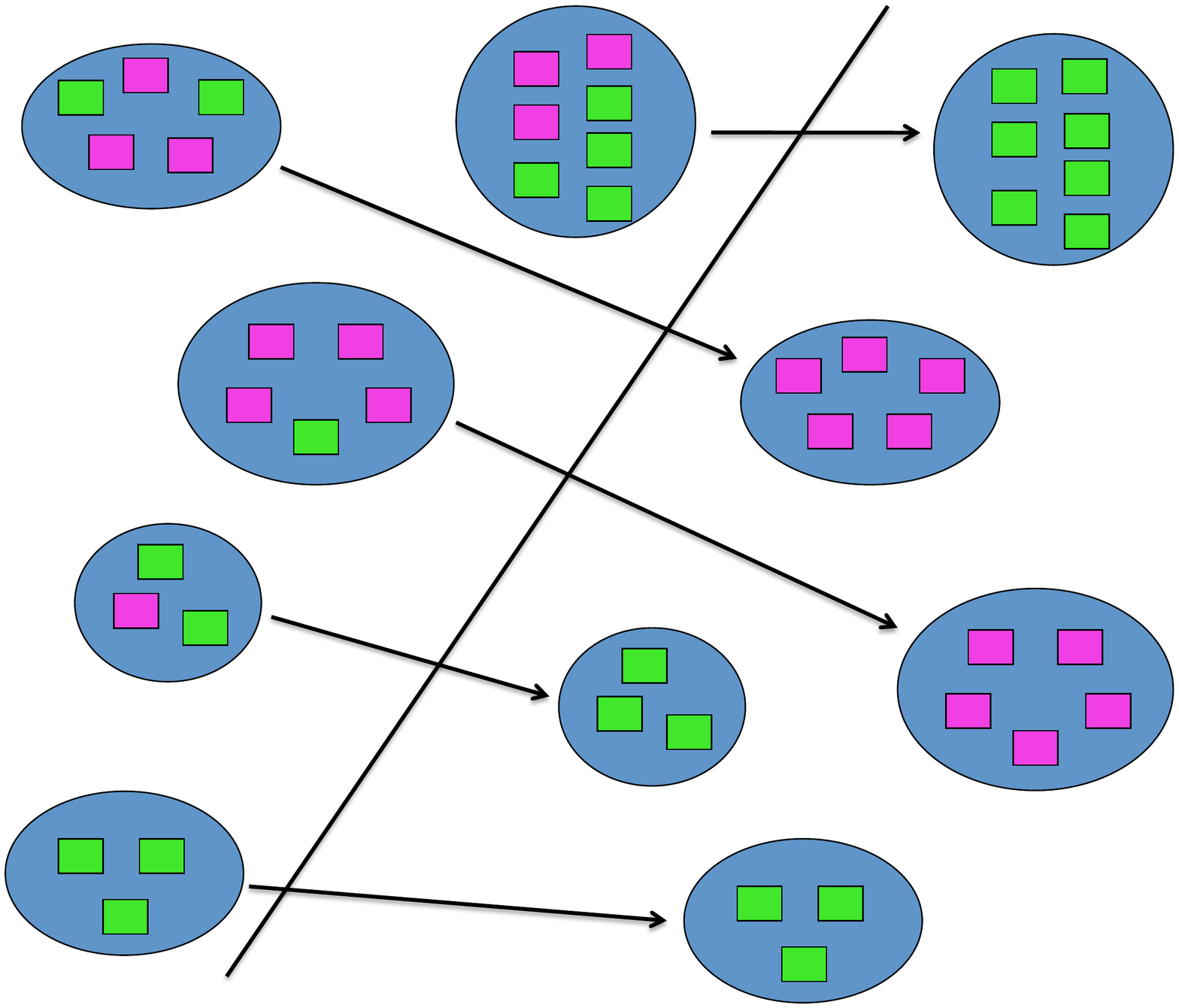}\\ \\ \\
\includegraphics[width=.50\textwidth]{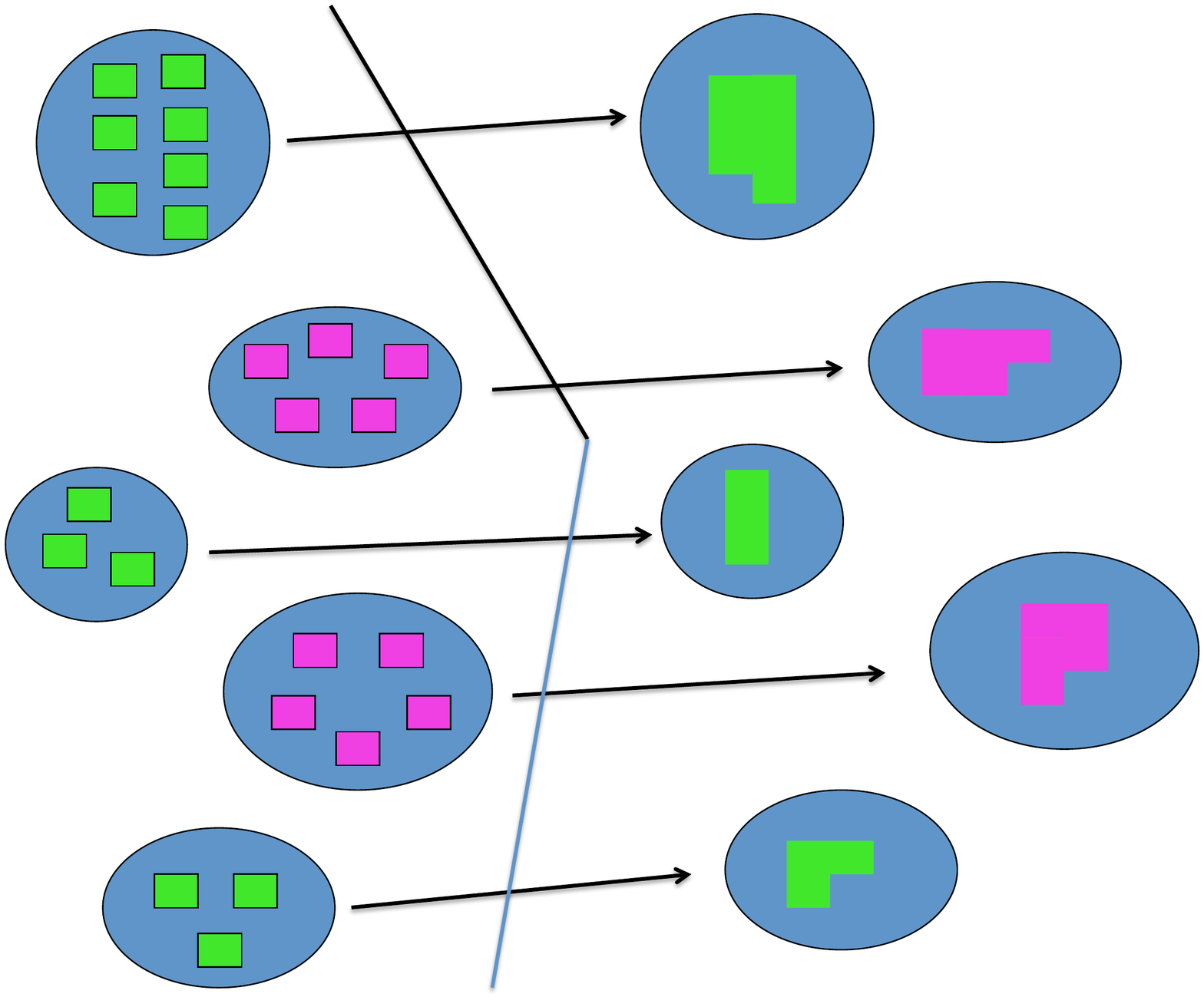}\quad
\includegraphics[width=.50\textwidth]{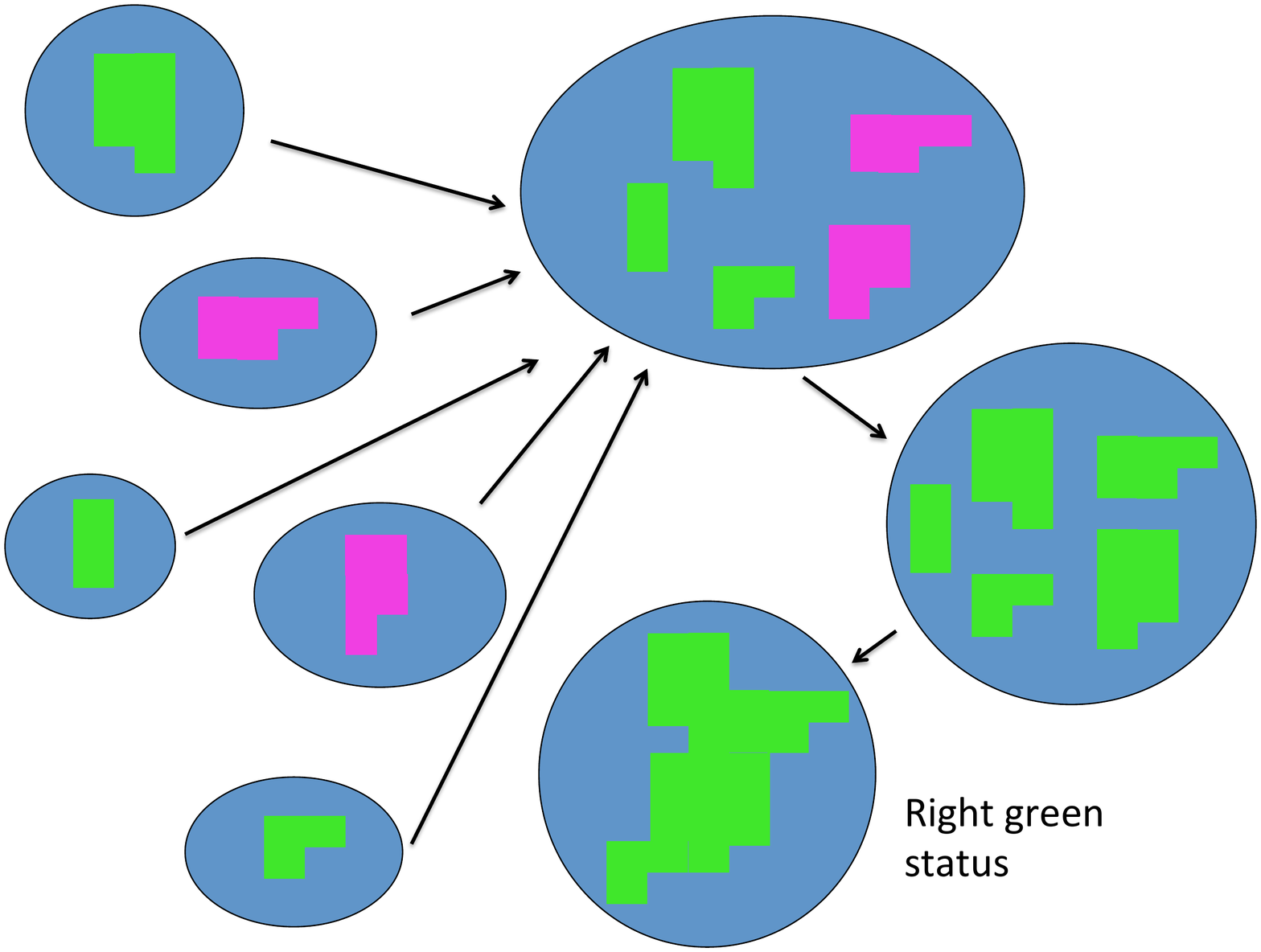}
\caption{Five aggregates are formed from the collection 23 initial items with 12 TF (green) and 11 TC (magenta), which means a NWT. The aggregates have respectively 3 (3 TF), 3 (2 T, 1 TC), 5 (2 TF, 3 TC), 5 (1 TF, 4 TC), 7 (4 TF, 3 TC) items. Majority rules yield 5 coarse-grained level-1  items with 3 TF and 2 TC. A second step coarse-grained from those five aggregates result in one TF large and unique item in adequacy with the hidden content of the collection of initial ground items. No uncertain aggregate has been treated.}
\label{puzzle-odd}
\end{figure}

Figure (\ref{puzzle-odd}) shows a treatment of all 23 items by forming 5 aggregates with respectively 3 (3 TF), 3 (2 T, 1 TC), 5 (2 TF, 3 TC), 5 (1 TF, 4 TC), 7 (4 TF, 3 TC) items. Applying majority rules to each aggregate yields 5 coarse-grained level-1  items with 3 TF and 2 TC. One additional coarse-grained of those five larger items gives one level-2  item, which embodies all the data with the label TF. The TF label is correct since 12 ground items are TF as seen in Figure (\ref{puzzle-odd}). No uncertain aggregate has been treated.

\begin{figure}
\includegraphics[width=.60\textwidth]{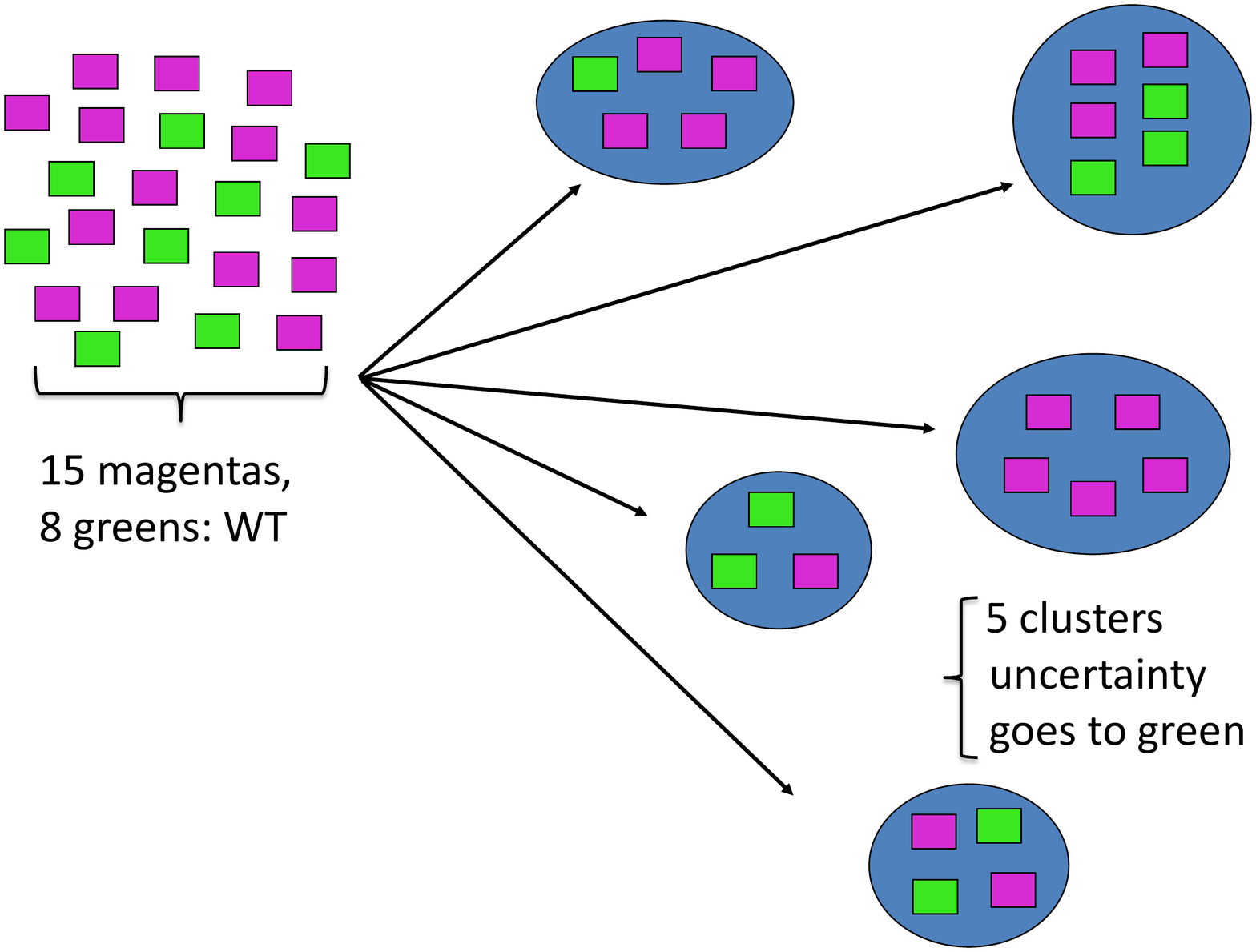}\quad
\includegraphics[width=.50\textwidth]{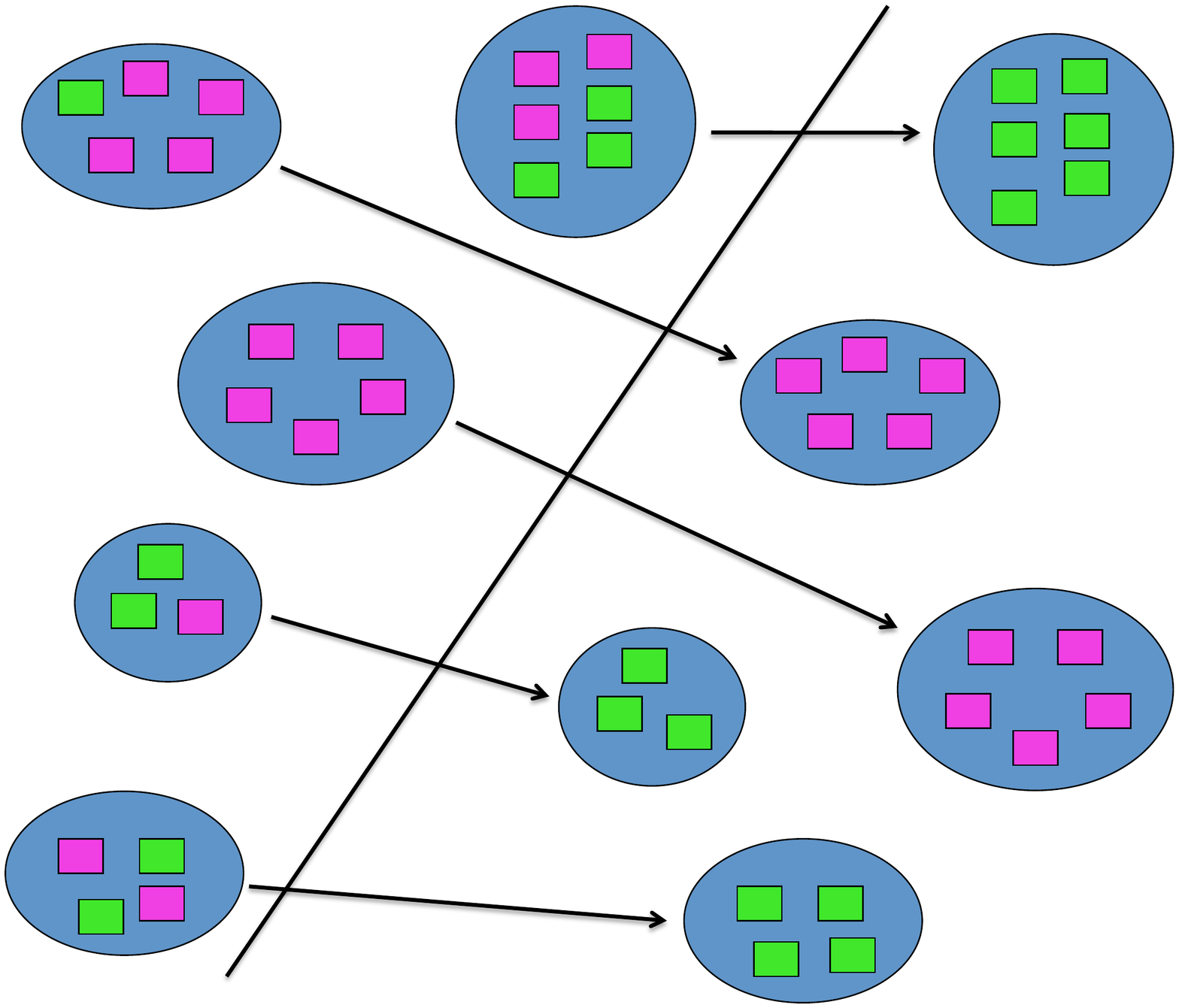}\\ \\ \\
\includegraphics[width=.50\textwidth]{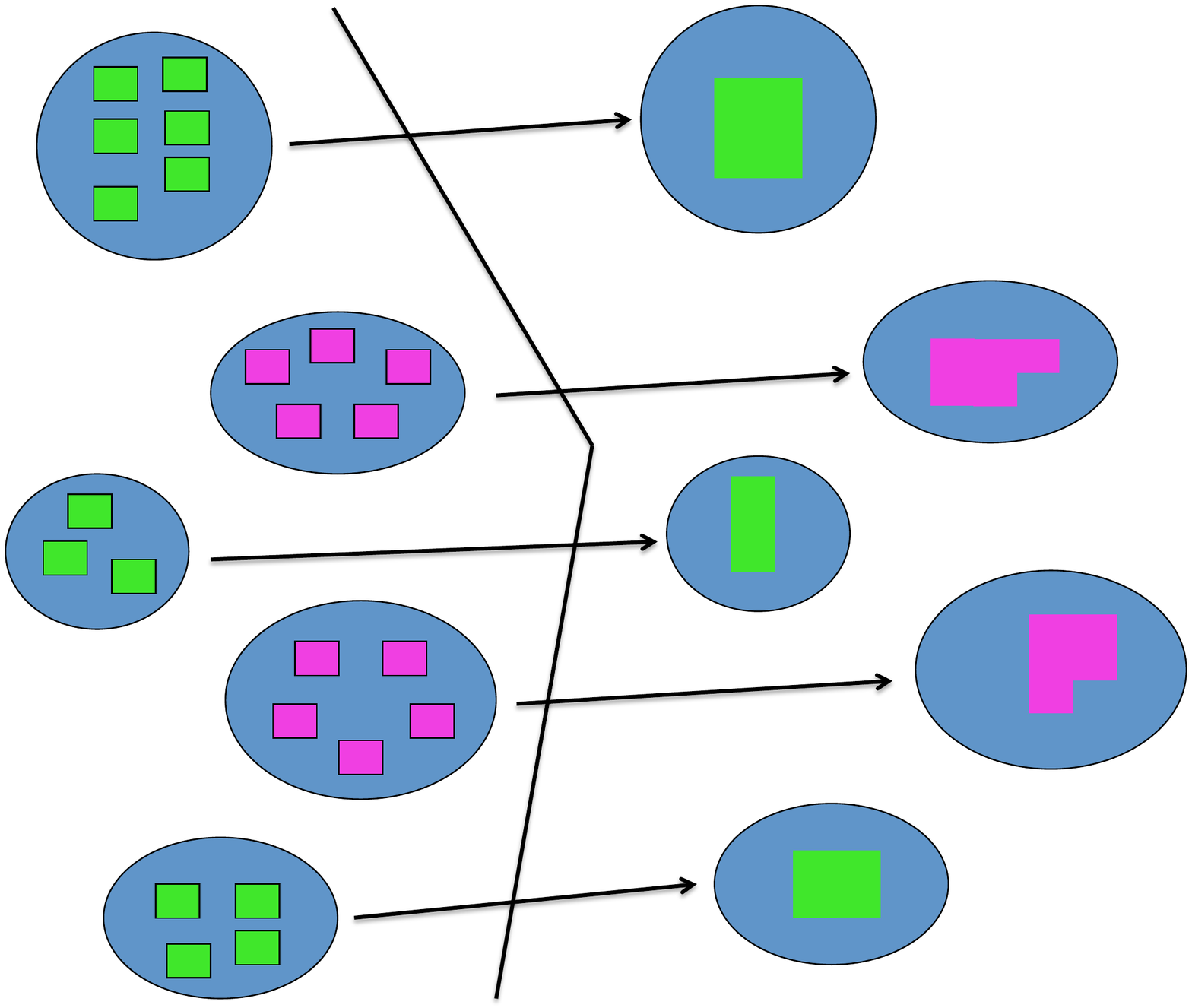}\quad
\includegraphics[width=.50\textwidth]{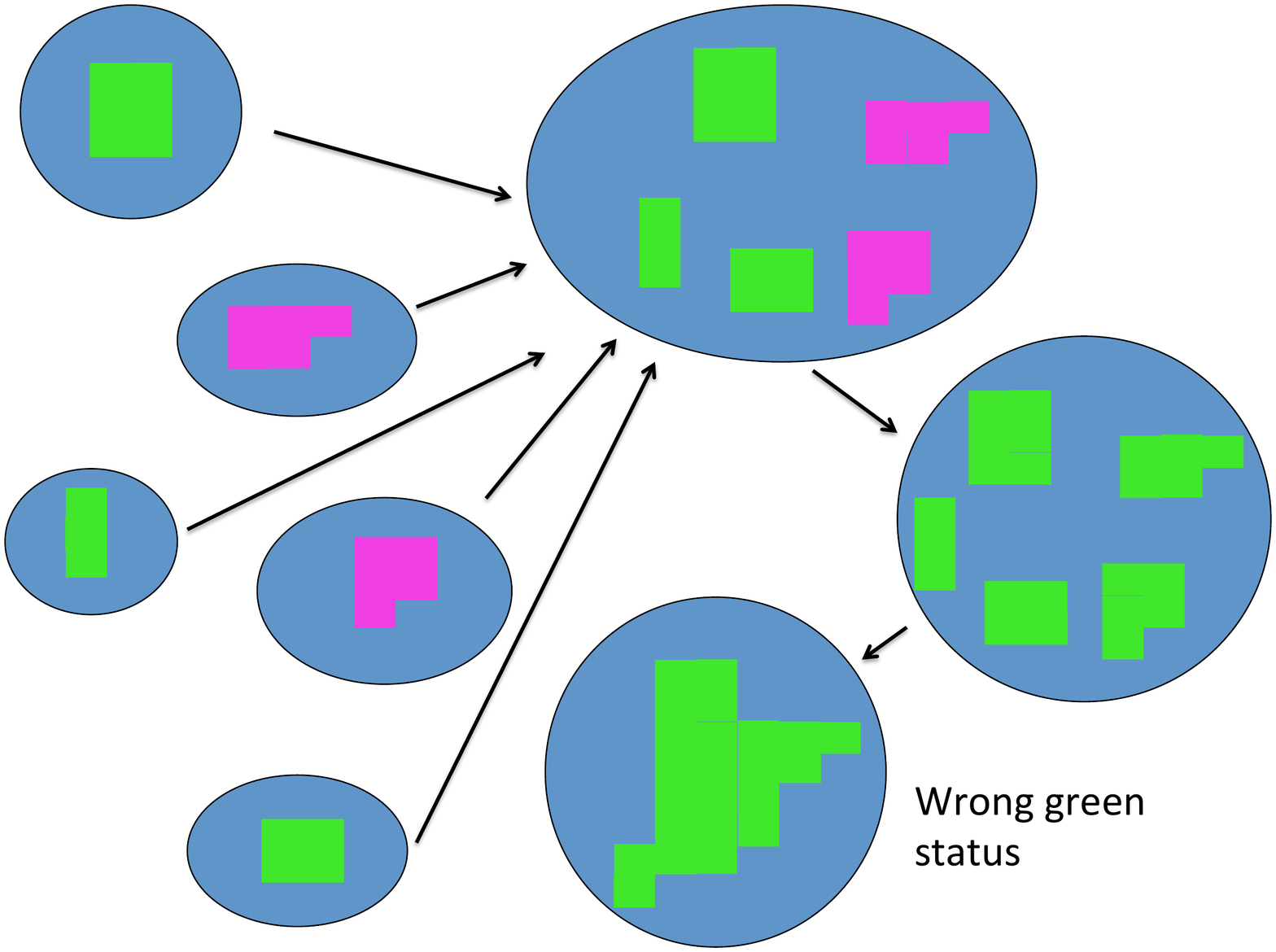}
\caption{(Five aggregates are formed with  3 (2 TF, 1 TC), 4 (2 TF, 2 TC), 5 (1 TF, 4 TC), 5 (0 TF, 5 TC), 6 (3 TF, 3 TC) items. Majority rules and uncertainty allocated to TF ($k=0)$ yield 3 TF (green) and 2 TC (magenta) level-1  items. The next coarse-grained yields a TF unique level-2  item, which is a wrong result.}
\label{puzzle-even}
\end{figure}

Figure (\ref{puzzle-even}) shows 23 items (15 TC, 8 TF) from which five aggregates are formed with 3 (2 TF, 1 TC), 4 (2 TF, 2 TC), 5 (1 TF, 4 TC), 5 (0 TF, 5 TC), 6 53 TF, 3 TC) items.  Majority rules and uncertainty allocated to TF ($k=0)$ yield 3 TF (green) and 2 TC (magenta) level-1 items. One additional coarse-grained of those five items yields a TF large and unique level-2  item with the wrong label (green). The related monitored person who is WT is left free of surveillance since labeled as NWT.

It is worth noting that if instead of applying the presumption of innocence to the two aggregates (2 TF, 2 TC) and (3 TF, 3 TC), the presumption of guilt is applied, the ending unique level-2 item get the right WT label.

Above cases were arbitrarily chosen only to illustrate the procedure of iterated treatment of aggregated items.

\section {Analytical solving of the procedure}

\subsection{The Equations}

To allow an analytically solving of the problem, I restrict all the grouping to four items. Starting with a density $p_0$ of TC ground items, the probability to collect a majority of TC items within a group of four writes,
\begin{equation}
p_{1}= p_0^4+4p_0^3(1-p_0)+6k p_0^2(1-p_0)^2 ,
\label{p1} 
\end{equation}
where first term corresponds to 4 TC, second term to (3 TC, 1 TF) and last term to (2 TC, 2 TF) which is allocated to TC with probability $k$.

Solving the fixed point Equation $p_1=p_0$ determines the dynamics driven by iteration of Eq.  (\ref{p1}). Three fixed points are obtained with two attractors $p_{WT}=1$, $p_{NWT}=0$ and a tipping point,
\begin{equation}
p_{c,k}=\frac{(1-6k)+\sqrt{13-36k+36k^2}} {6(1-2k)},
\label{k4} 
\end{equation}
which gives $p_{c,0}=\frac{1+\sqrt{13}} {6} \approx 0.77$,  $p_{c,1/2}=\frac{1}{2}$ and $p_{c,1} = \frac{5-\sqrt{13}} {6} \approx 0.23$ for respectively $k=0, \frac{1}{2}, 1$. Therefore $0\leq k \leq \frac{1}{2} \Rightarrow 0.23\  \leq p_{c,k}\leq \frac{1}{2}$  and $\frac{1}{2}\leq  k \leq 1 \Rightarrow \frac{1}{2} \leq p_{c,k}\leq  0.77$.

For a collection of $N$ ground items, the number $n$ of necessary successive coarse-grained iterations to get the unique n-level item, which encompasses all of them is,
\begin {equation}
n=\frac{ \ln N} {\ln 4} .
\label{n4} 
\end{equation}

However, to get the single giant item starting from a proportion $p_0$ of TC ground items, does not ensures to reach one of the two attractors. To make sure an attractor is reached requires a number $m$ of iterations given by \cite{geo},
\begin {equation}
m=I\Bigg[\frac{1}{\ln \lambda_k} \ln \frac{1}{\vert 1-\frac{p_0}{p_{c,k}}\vert} \Bigg] +2 ,
\label{m} 
\end{equation}
where $I[\dots]$ means taking the integer part of $[\dots]$ and $ \lambda_k=\frac{dp_1}{dp_0} \vert_{p_{c,k}}$ with $ \lambda_0= \lambda_1\approx 1.64$ and $\lambda_{\frac{1}{2}}=\frac{3}{2}$. 

\begin{figure}
\includegraphics[width=.60\textwidth]{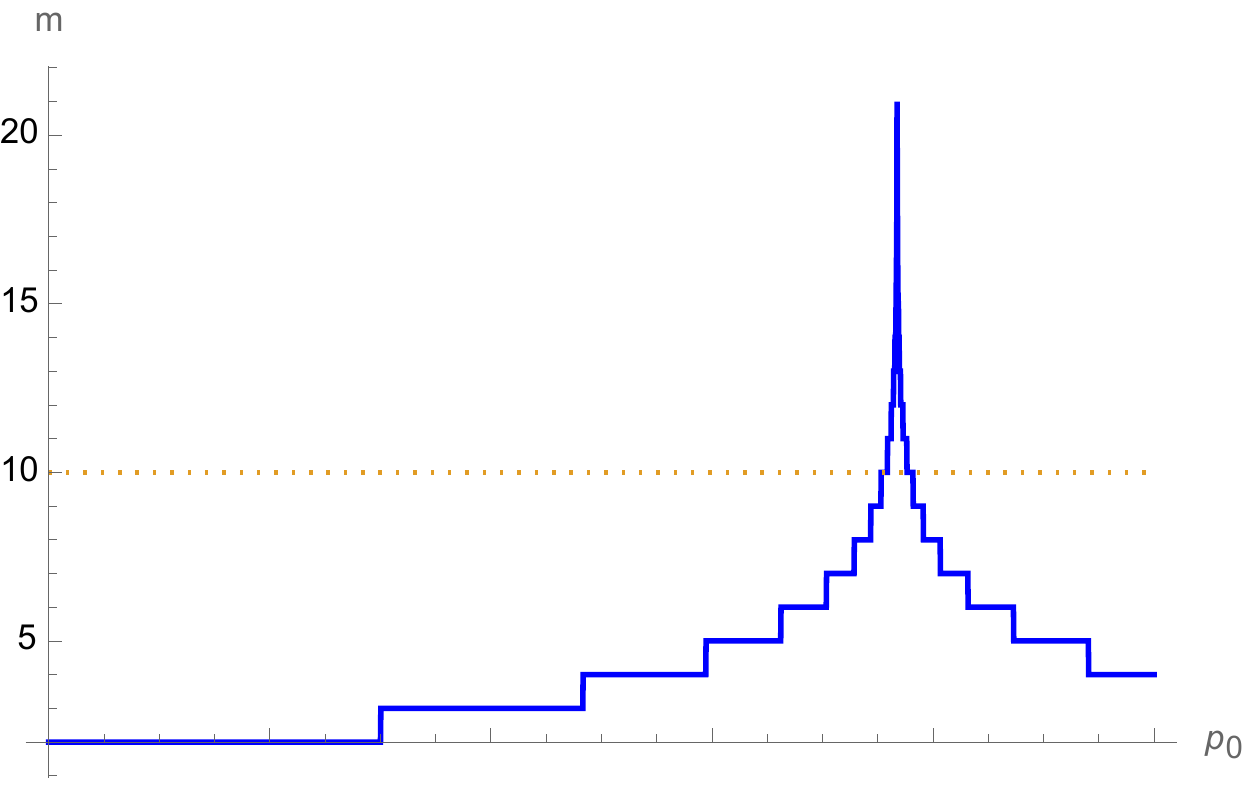}\
\caption{Number of iterations required to reach one of the two attractors and thus get a deterministic label of the giant item as a function of the proportion $p_0$ of ground items.}
\label{mp}
\end{figure}

Eq. (\ref{m}) is an approximate value \cite{geo}. Nevertheless, Figure (\ref{mp}) shows that than all its values are less than 10 for most ranges of $p_0$ besides a cusp reaching 20 in the immediate vicinity of $p_0\approx 0.77$. There, the required amount of data is out of reach with $N=4^{20}$ (Eq. (\ref{n4})).

Moreover, when $n<m$, the value $p_n$ may not be 0 or 1 and the label of the giant item is probabilistic.  Nevertheless, this feature must be mitigated by noting that in the very close vicinity of the attractor, dealing with real numbers, $p_n\approx$ 0 or 1. Indeed, in most cases, one of the attractor is reached within $n$ iterations or less assuming assuming a reasonable accuracy. 

\subsection{An ineradicable wrong labelling}

While $p_{WT}$ and $p_{NWT}$ are stable attractors, $p_{c,k}$ is a tipping point. Starting from $p_0<p_{c,k} \Rightarrow p_0>p_1>p_2>...>p_m \approx p_{NWT}=0$ while $p_0>p_{c,k}$ $\Rightarrow p_0<p_1<p_2<...<p_m \approx p_{WT}=1$ with $m$ given by Eq.  (\ref{m}).

Given that the  threshold to be a WT is at $0.50$, I can now associate a validity status for the labelling of the final unique item as a function of $p_0$ and $p_{c,k}$. When $k=0$ the iterated coarse-grained scheme yields the right label for $0\leq p_0\leq 0.23$ and $0.50 \leq p_0\leq 1$. A wrong status is obtained when $0.50 \leq p_0\leq 0.77$.

\begin{figure}
\centering
\includegraphics[width=1\textwidth]{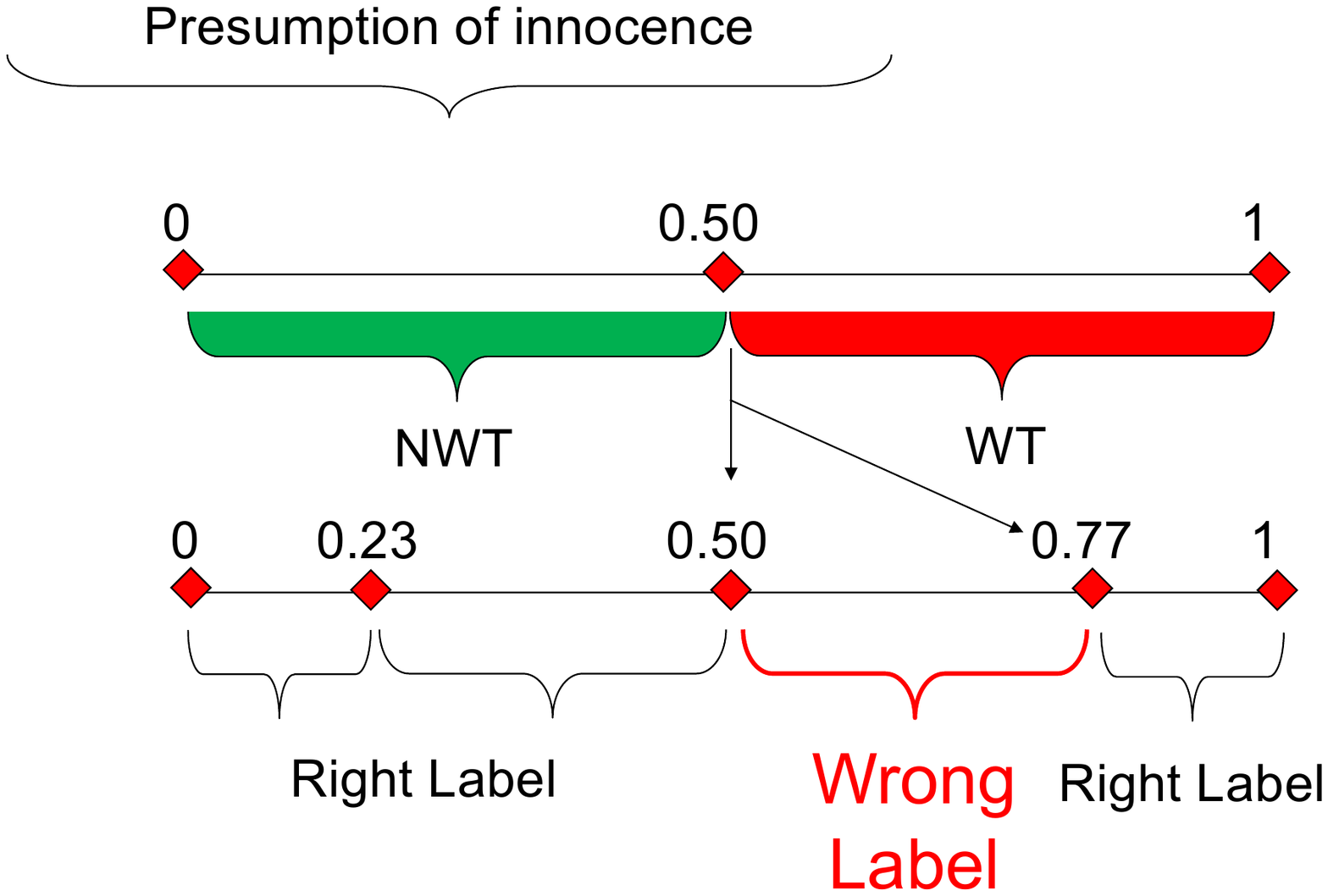}
\caption{Outcomes of status for the large and unique item using $k=0$. The right status is obtained for $0\leq p_0\leq 0.50$ (NWT) and $0.77 \leq p_0\leq 1$ (WT) but a wrong status (NWT)is found when $0.50 \leq p_0\leq 0.77$.}
\label{ag1}
\end{figure}

\begin{figure}
\centering
\includegraphics[width=1\textwidth]{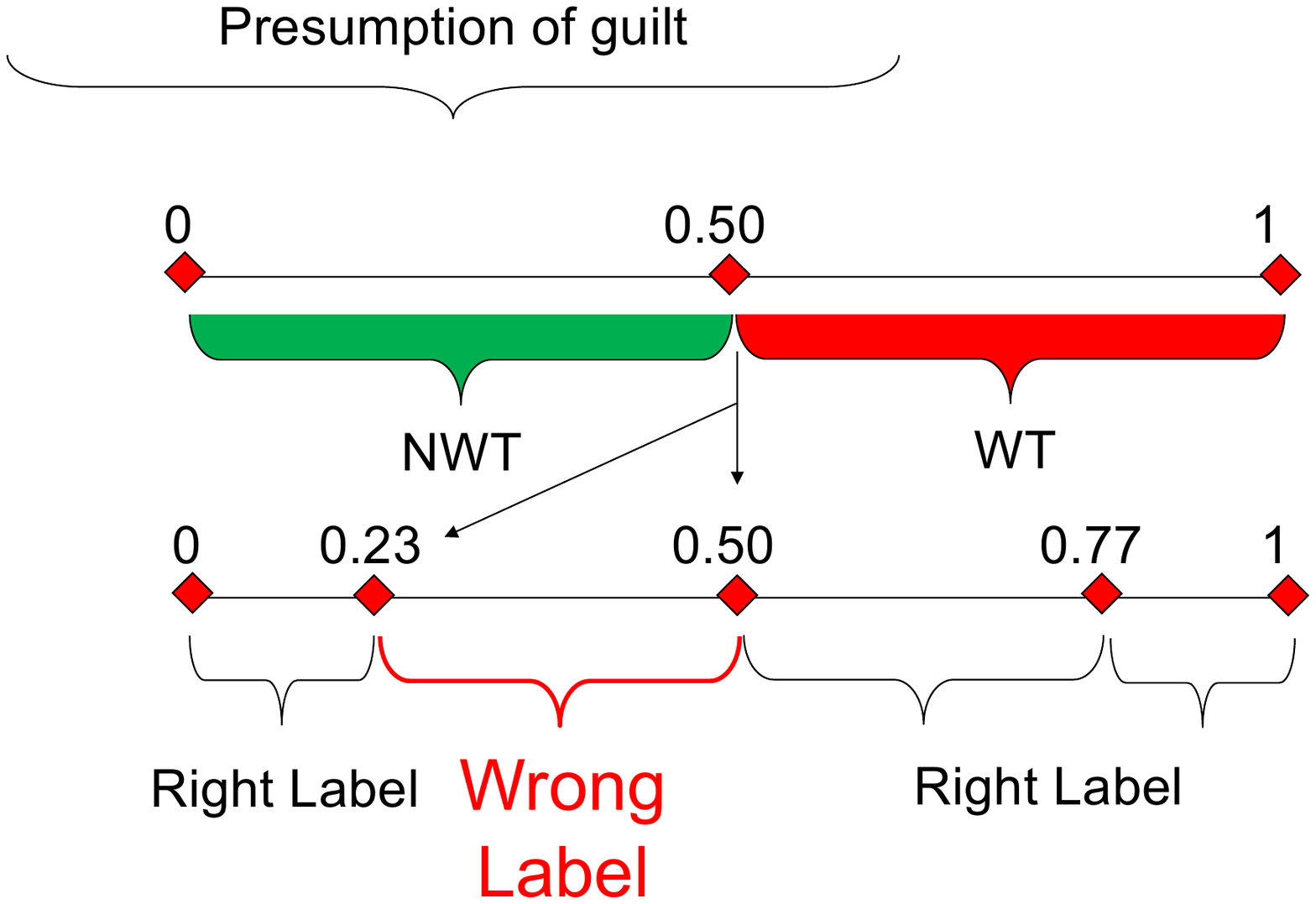}
\caption{Outcomes of status of the large and unique item using $k=1$. The right status is obtained for $0\leq p_0\leq 0.23$ (NWT) and $0.50 \leq p_0\leq 1$ (WT) but a wrong status (WT) is found when $0.23 \leq p_0\leq 0.50$.}
\label{ag2}
\end{figure}

In contrast, when $k=1$, the right label is obtained for $0\leq p_0\leq 0.23$ and $0.50 \leq p_0\leq 1$ with a wrong label for $0.23 \leq p_0\leq 0.50$. The summary is shown in Figures (\ref{ag1}, \ref{ag2}).

The analysis of the mirror model has revealed that the coarse-grained treatment yields the exact outcome for low and large proportions of TC ground items. However, for range around fifty percent of TC items, the outcome is systematically wrong either at the benefit of WT (presumption of innocence) or at the expense of NWT (presumption of innocence), depending on the presumption applied locally to tag uncertain aggregates of items. In the first case, some WT are stamped NWT but no NWT is wrongly tagged WT. In the second case, it is the other way around, no NWT is missed, but some NWT are mislabeled as WT.

\subsection{Combining different sizes of aggregates}

Above analysis was restricted to aggregates with 4 items and the instrumental values 0.23 and 0.77 determining the correctness status obtained for the giant element shown in the Figures (\ref{ag1}, \ref{ag2}), depend on this size. To check the robustness of the results, I now extend the coarse-grained scheme by considering a combination of different sizes $r$ of aggregates with the respective proportions $g_r$ satisfying,
\begin{equation}
\sum_{r=1}^{r=R}g_r=1,
\label{gr} 
\end{equation}
where $R$ is the larger size allowed. Eq. (\ref{p1}) is then replaced by,
\begin{equation}
p_{R,1}=\sum_{r=1}^{R} \left\{  g_{r} \left [ \sum_{l=\bar r+1}^{r}   {r \choose l} p_{0}^l  (1-p_0)^{r-l} +\delta [\bar r-\frac{r}{2}]k {r \choose r/2}  p_{0}^{\frac{r}{2}}  (1-p_{0})^{\frac{r}{2}} \right ] \right\}  \ ,
\label{Tr} 
\end{equation}
where $\bar r \equiv I [\frac{r}{2}]$ with $I [...]$ meaning integer part of $[...]$ and $\delta [x]$ is the Kronecker function. 

While Eq. (\ref{Tr}) yields the same attractors $p_{WT}=1$ and $p_{NT}=0$ as for $r=4$, the threshold $p_{c,R,k}$ can now be located anywhere between 0 and 1 as a function of $\{g_r\}$ and $R$. Figures (\ref{ag1}) and (\ref{ag2})  are thus still valid but with different values instead 0.23 and 0.77. Those values must be calculated numerically from the fixed point equation $p_{R,1}=p_0$.

Large values of $r$ reduce the probabilities of a tie and thus reduce the distance of $p_{c,R,k}$  from $0.50$. But in contrast, the value $r=2$ places the threshold $p_{c,R,k}$ far away from 0.50 at 0 (for $k=0$) or 1 (for $k=1$). For instance $r=6,8,10$ yield respectively  $p_{c,0}=0.65,0.61,0.58$ and $p_{c,1}=0.35,0.39,0.42$.

For instance, a combination of sizes with $g_1=0.20,g_2=0.30,g_3=0.20,g_4=0.20,g_5=0.10$ yields $p_{c,R,0}=0.85$ and $p_{c,R,1}=0.15$. The case $p_{c,R,0}=0.85$ gives the series $p_0=0.70, p_{R,1}=0.66, p_{R,2}=0.60, p_{R,3}=0.52, p_{R,4}=0.41, p_{R,5}=0.28, p_{R,6}=0.15, p_{R,7}=0.05, p_{R,8}=0.01, p_{R,9}=0.00$. Nine successive coarse-grained tag a WT wrongly as NWT.

Accordingly, the existence of ranges $0.50<p_0<p_{c,rR0}$  and $p_{c,R,1}<p_0<0.50$ for which WT are diagnosed as NWT and NWT people are labeled WT is a solid result. Moreover while these ranges can be modified, they cannot be shrunk to zero.

\section{Difference between the Mirror model and both Galam models of voting and opinion dynamics}

The Mirror model, the Galam hierarchical voting model and the Galam opinion dynamics model deal respectively with different realities. However, all three models are based on local updates and use identical equations. But the respective "physics" of the variables and parameters is specific to each model, as are the implications of the results obtained. 

\subsection{Galam bottom-up voting model}

The model describes a bottom-up voting hierarchy. It starts from a bottom level made up of groups of $r$ people, each one electing a representative using local majority rules \cite{geo}. In case of a tie for even size groups, the representative is elected along the incumbent president. Those elected representatives constitute the level-1 of the hierarchy. They in turn constitute groups of $r$ members to elect representatives still following  local majority rules and tie breaking at the benefit of the incumbent president. Those elected representatives constitute higher level-2 of the hierarchy.

The same election process is iterated again and again with at each higher level a number of elected representatives divided by $r$. At some level $n-1$, one single group of representatives elects the president of the hierarchy at level $n$..

\begin{figure}
\centering
\includegraphics[width=1\textwidth]{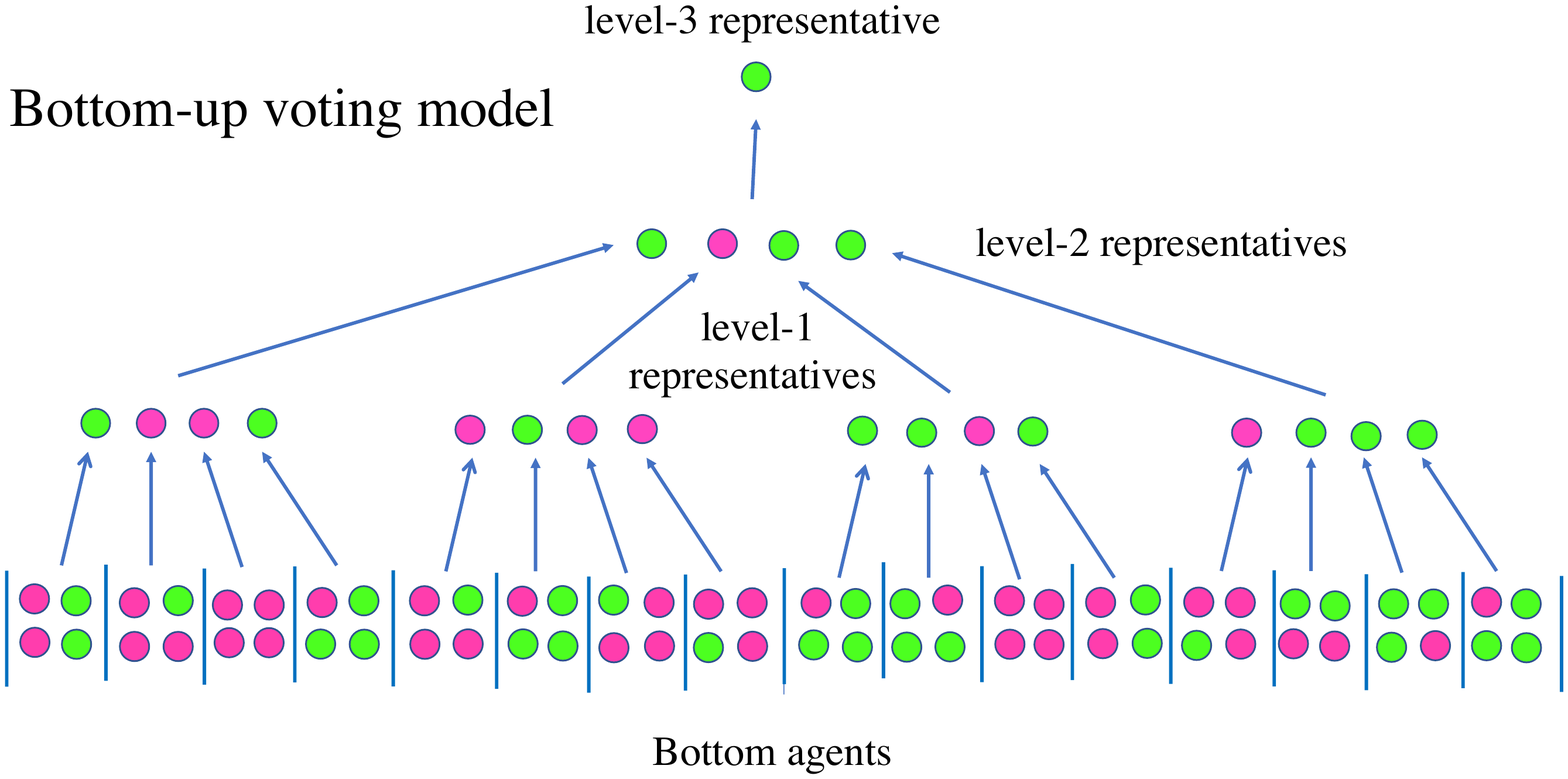}
\caption{A bottom-up voting hierarchy with 3 levels, 64 agents at the bottom, voting groups of four people, and a total of 85 agents. The agents are respective supporters of two competing parties represented by the colors green (G) and magenta (M).  In this case, the incumbent president is G. At the bottom there are 35 M agents and 29 G agents, who elect 16 level-1 representatives with 9 G and 7 M. At level-2, 3 G and 1 M have been elected, who in turn elected a G president at level-3. }
\label{M1}
\end{figure}

A hierarchy of $n$ levels requires $r^n$ agents at the bottom and involves $r^{n-1}, \dots, r, 1$ representatives at respective levels $1, 2, \dots, n-1, n$. An illustration is given in Figure  (\ref{M1}) for a hierarchy with 3 levels, voting groups of four people, a bottom with 64 agents and a total number of 85 agents.

\subsection{Galam opinion dynamics model}

The model describes the dynamics of opinion among a community of agents where individuals shift opinion via local and informal discussions among groups of $r$ agents \cite{glob}. In each discussing group, all agents update their opinion by adopting the opinion shared by the local majority. In case of a tie for an even size group, the opinion is selected according to the shared prejudice among the $r$ agents. Afterwards, groups are dispersed and all agents are reshuffled. This three-step scheme is then iterated till reaching one of the attractor of the dynamics. 

The number $n$ of updates required to reach an attractor plays the same role as the number of levels in the bottom-up voting hierarchy. However, here the same agents keep shifting opinions while in the voting model no agent changes opinion with more agents being added at each additional elective level. 

\begin{figure}
\centering
\includegraphics[width=1\textwidth]{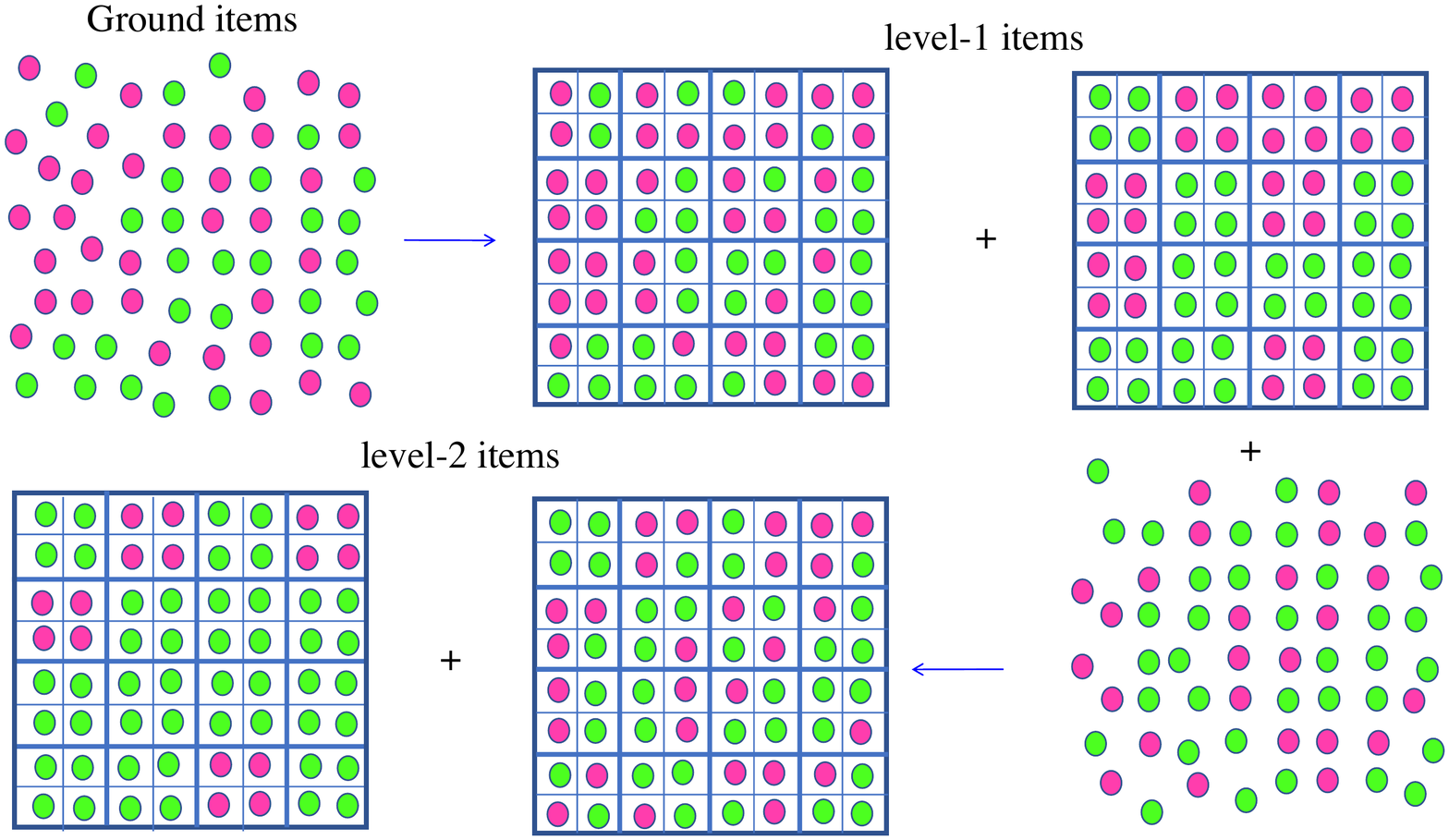}
\includegraphics[width=1\textwidth]{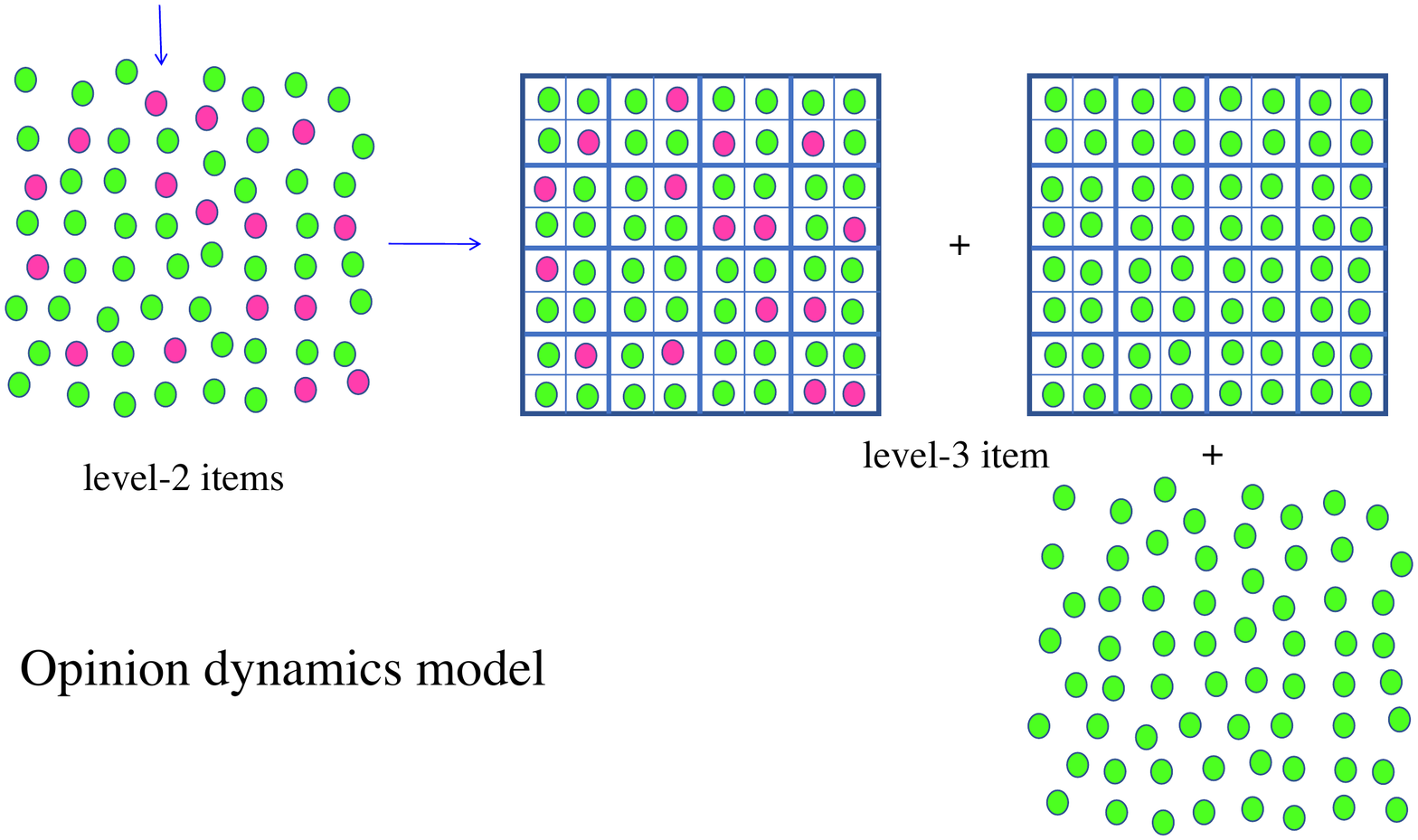}
\caption{A population of 64 agents with update groups of four agents and three iterated updates. Two choices respectively green (G) and magenta (M) are competing with an initial support of 35 M agents and 29 G agents. At a tie the prejudice favors G. The first update yields 28 M and 36 G. A second update gives 16 M and 48 G. Last update turns G every one.}
\label{M2}
\end{figure}

An illustration is given in Figure  (\ref{M2}) for a population of 64 agents with update groups of four agents and three iterated updates.

The model can also include heterogeneous agents with three types of psychological traits, which are flexible agents or floaters who do hold an opinion but update it locally following the local majority rule, inflexible, stubborn or committed agents who hold an opinion and stick to it whatever is the local majority, contrarian agents who hold an opinion but shifts it against the local majority.

\subsection{Data mirror model}

Contrary to the other models, which deals with a collection of agents, the data mirror model deals with one single person. The procedure processes a set of data collected during the monitoring of this person. Each element is labeled either TC or TF. The ground items are then gathered in groups of $r$ to be synthesized into larger pieces. The process is then iterated in a coarse-grained pattern until it encompasses all the ground items.

The number $n$ of iterations is the equivalent of the number of hierarchical levels and the number of opinion updates in above tow models.

\begin{figure}
\centering
\includegraphics[width=1\textwidth]{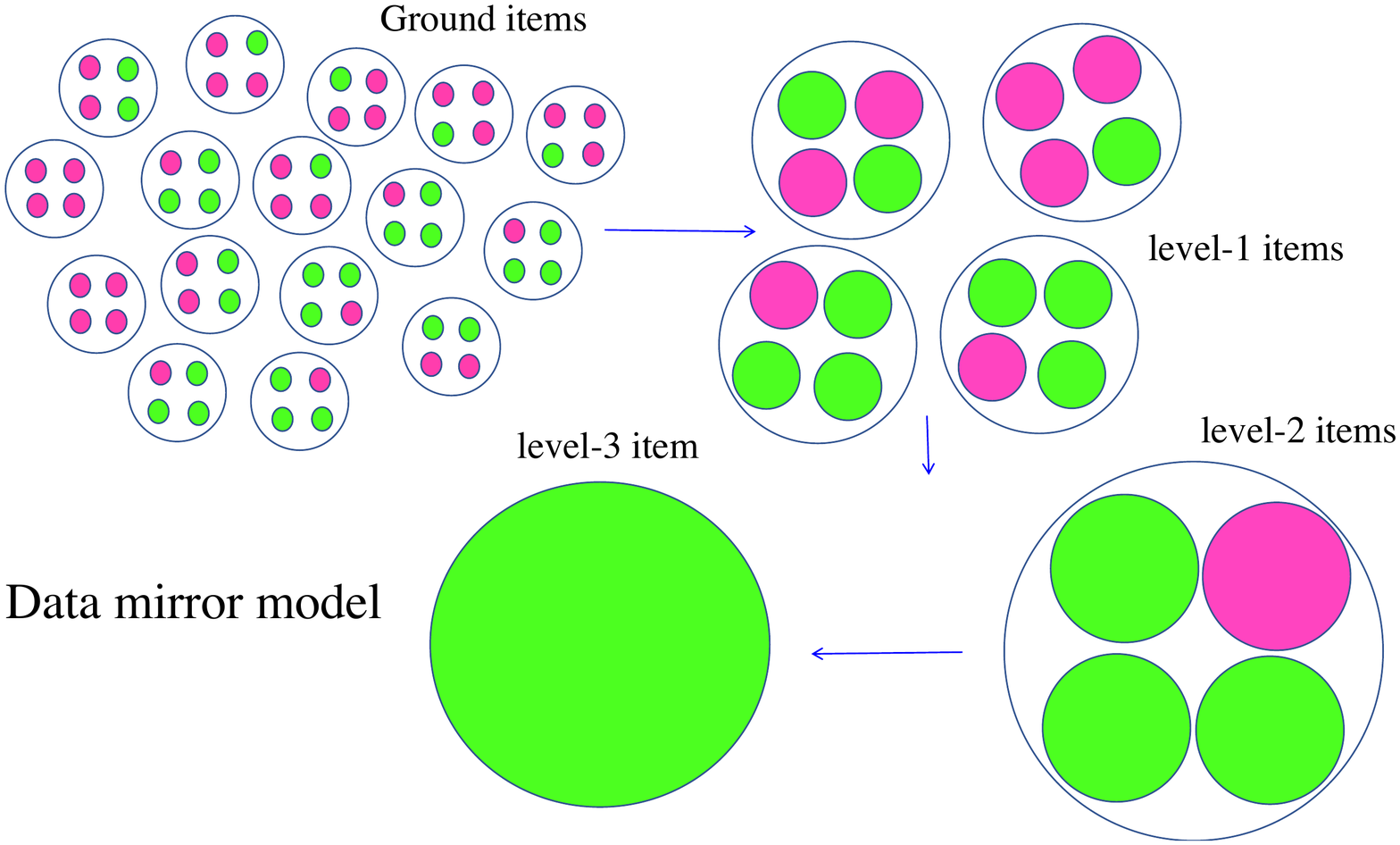}
\caption{A collection of 64 ground items with 35 TC versus 29 TF. Three successive aggregations by groups of four items lead to one level-3 item labeled TF.}
\label{M3}
\end{figure}

An illustration is given in Figure  (\ref{M3}) for a collection of 64 ground items with 35 TC items versus 29 TF items. Three successive aggregations by groups of four items, leads to one level-3 item labeled TF. The round shapes of the items are only used as visual support and have no particular meaning.

\section{A large scale simulation}

To illustrate the iterated coarse-grained scheme I ran a simulation with a sample of $4^8=65536$ ground items, which requires $8$ iterations to encompass all ground items. The proportion of TC items is $p_0=0.750$. To make easy the visualization of the iterated scheme, the ground items are distributed randomly on a square lattice  $256 \times 256$ as shown in Figure Figure (\ref{f0}). This choice of a square lattice has no meaning in itself beside being an easy visual support. Having $p_0=0.750$ means that the related person is a WT. 

To implement the processing of data, items are aggregated by groups of 4 nearest neighbors. Iterations are implemented 
by applying the presumption of innocence at uncertain aggregates, i.e. are labelled TF. TC items are in red, TF items in green, uncertain aggregates in blue before they are turned TF in green.

\begin{figure}
\centering
\includegraphics[width=.6\textwidth]{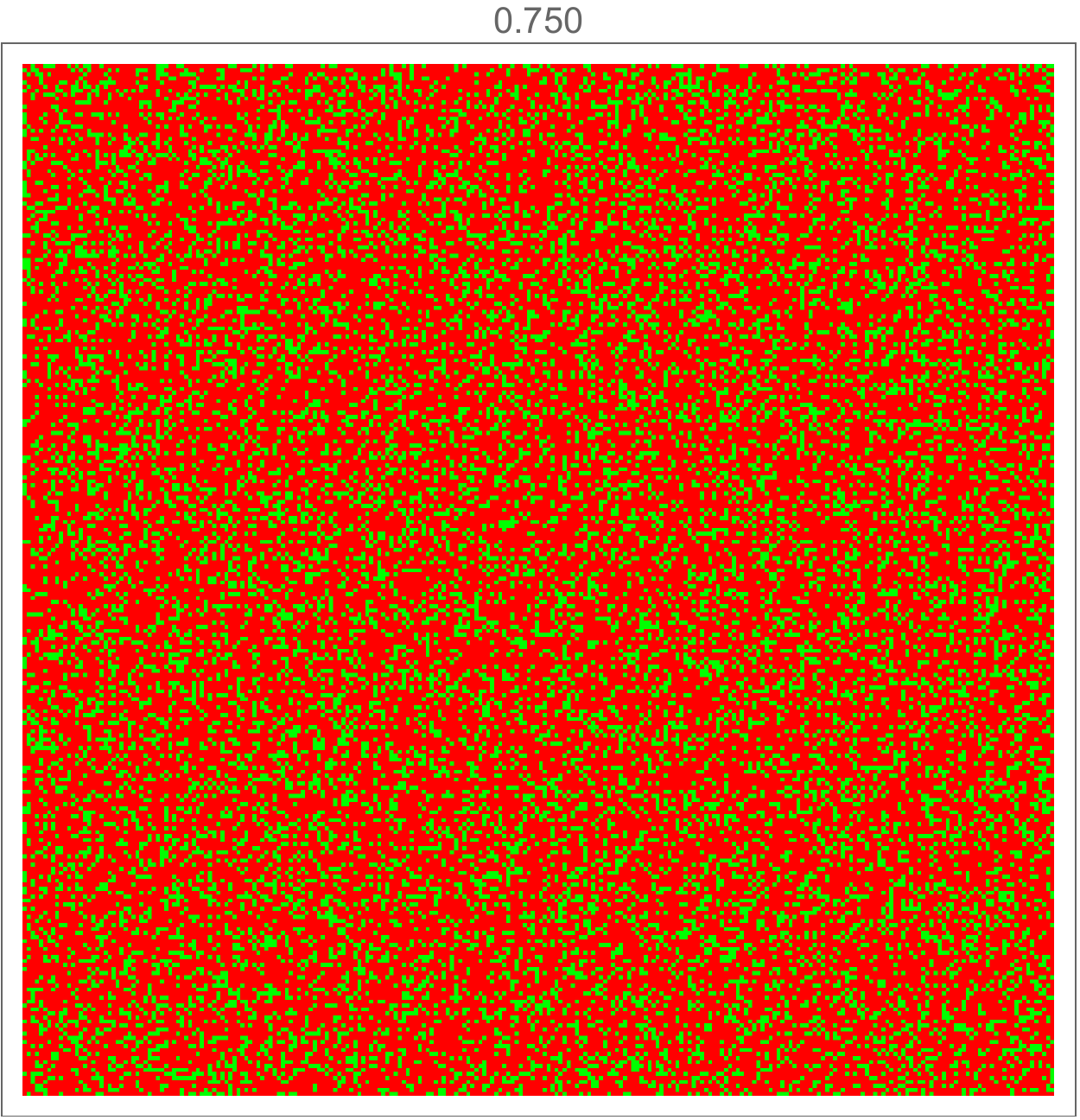}\quad
\includegraphics[width=.4\textwidth]{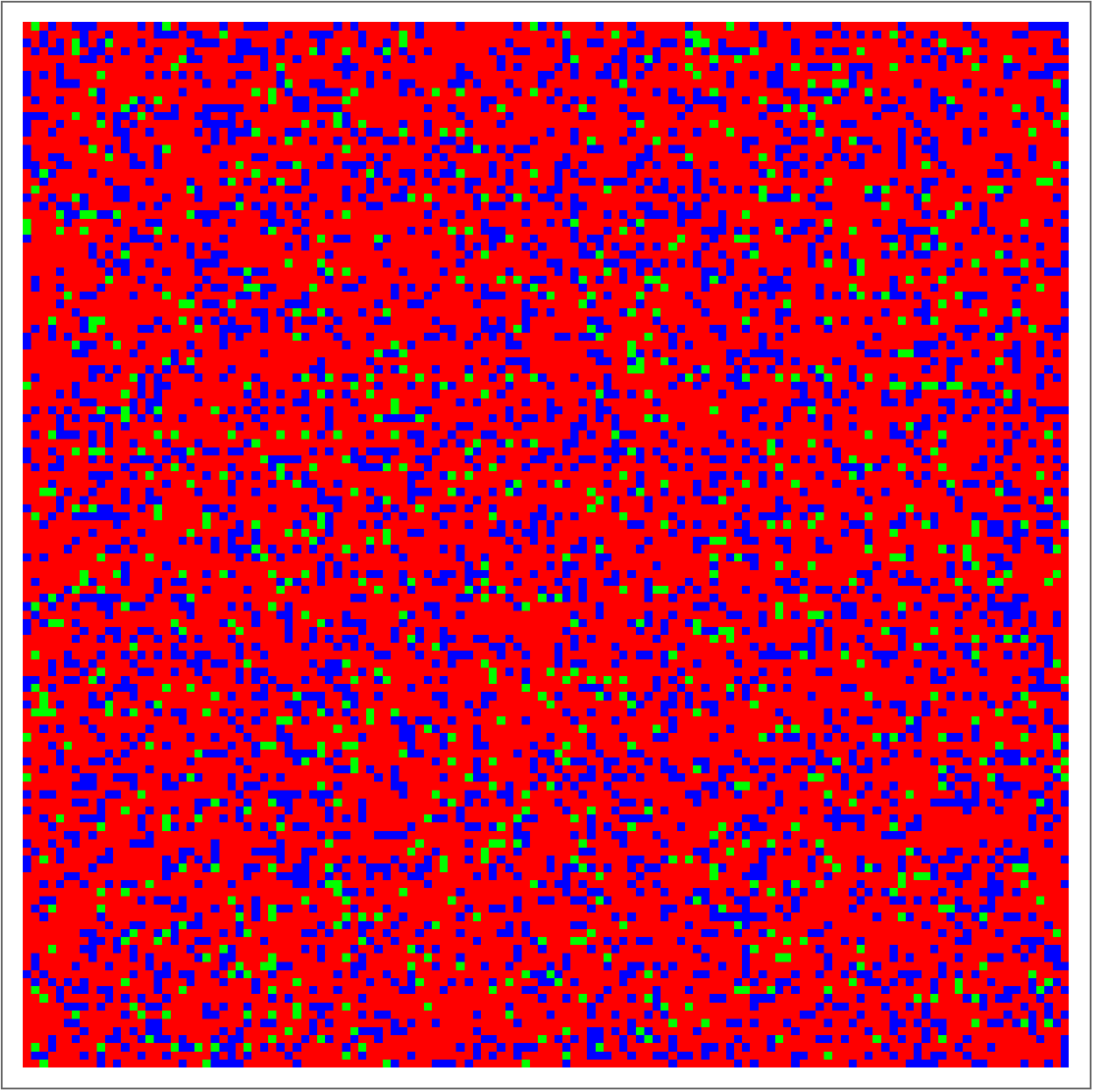}\quad
\includegraphics[width=.4\textwidth]{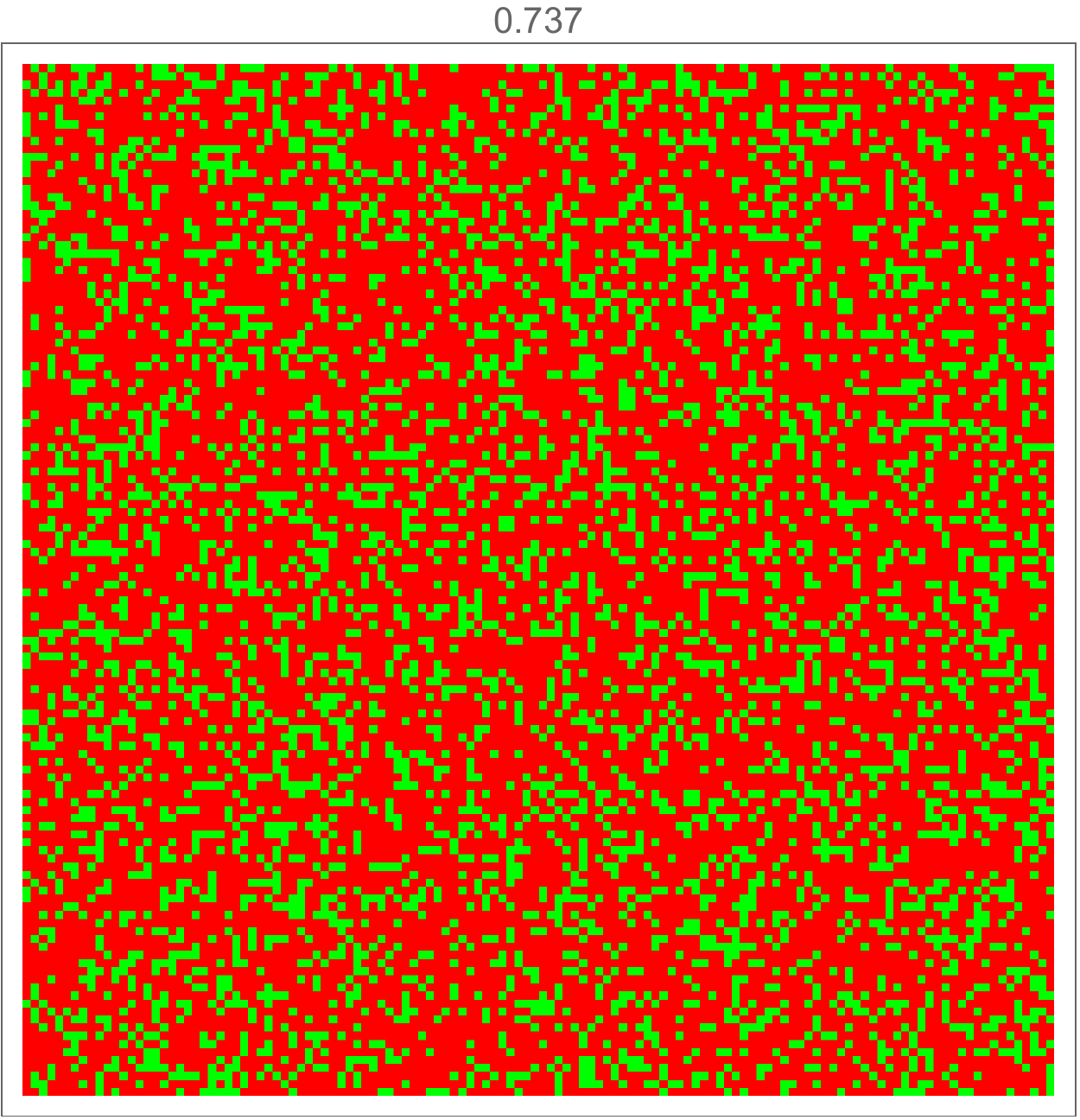}\quad
\includegraphics[width=.4\textwidth]{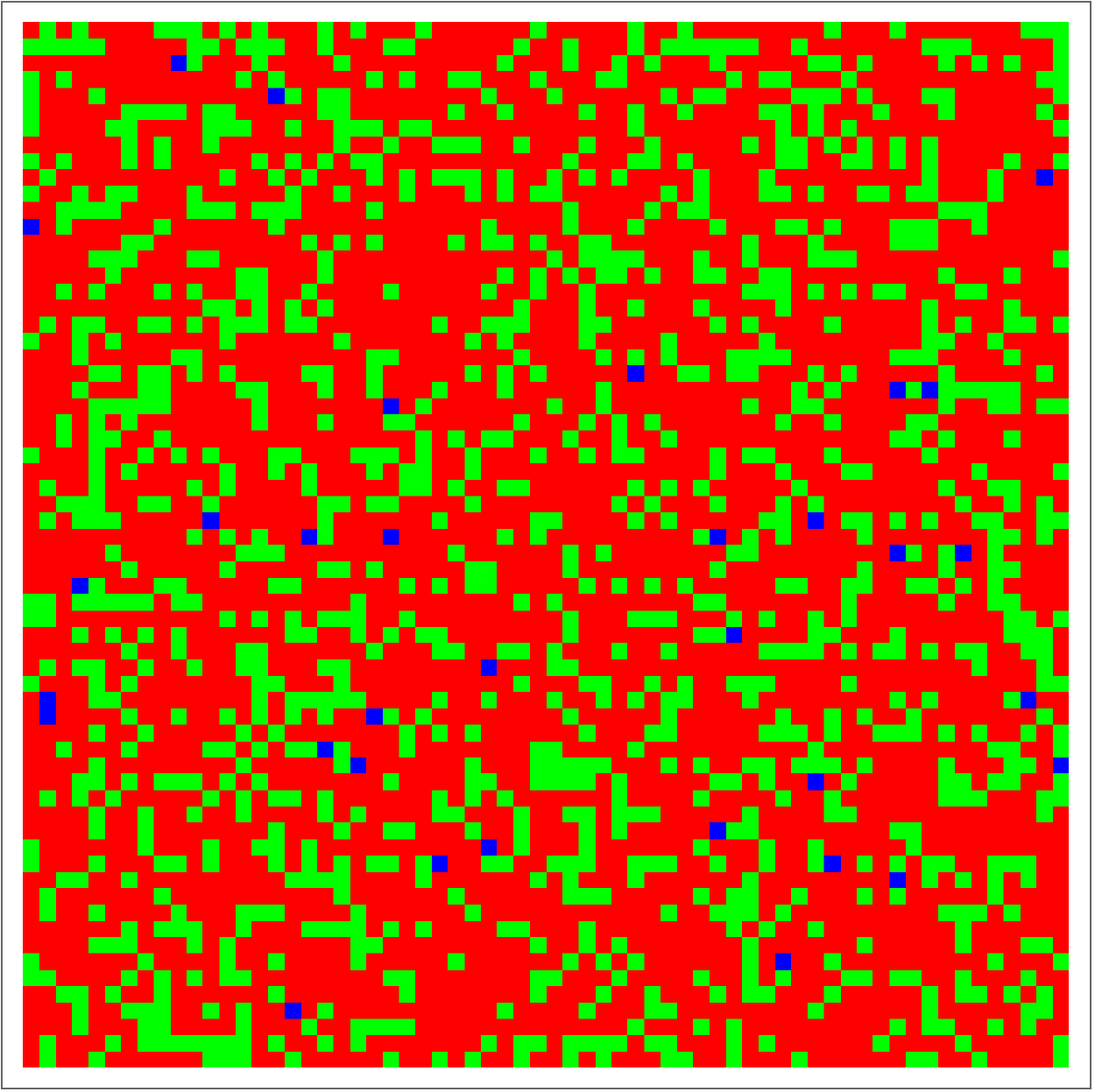}\quad
\includegraphics[width=.4\textwidth]{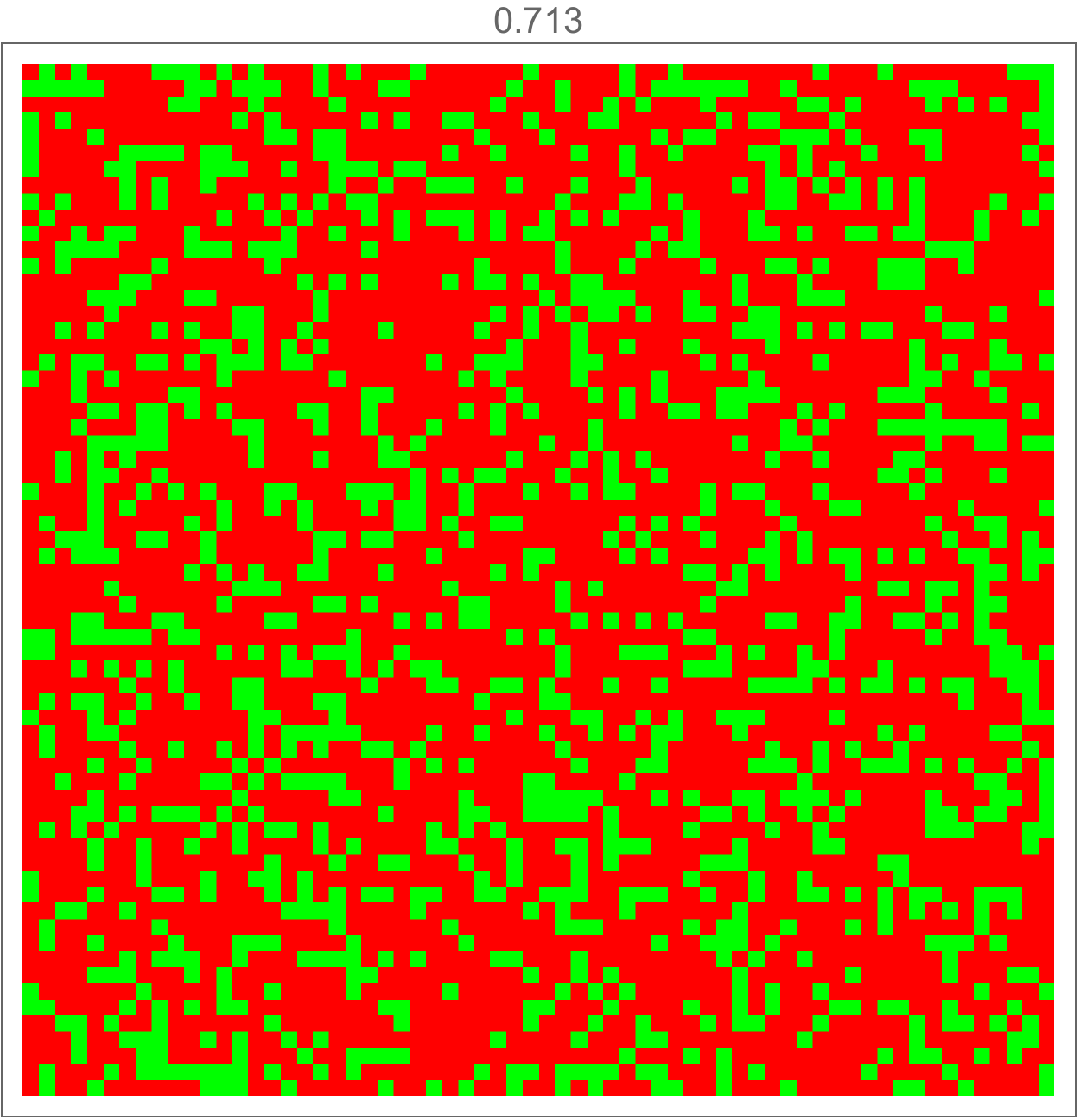}\quad
\caption{$N=4^8=65536$ ground items with $p_0 = 0.750$ TC items are shown at the top. The related person is thus a WT. The first iteration is shown in the middle with $p_1 = 0.737$ and the second iteration is exhibited at the bottom with $p_2 = 0.713$.}
\label{f012}
\end{figure} 

\begin{figure}
\centering
\includegraphics[width=.4\textwidth]{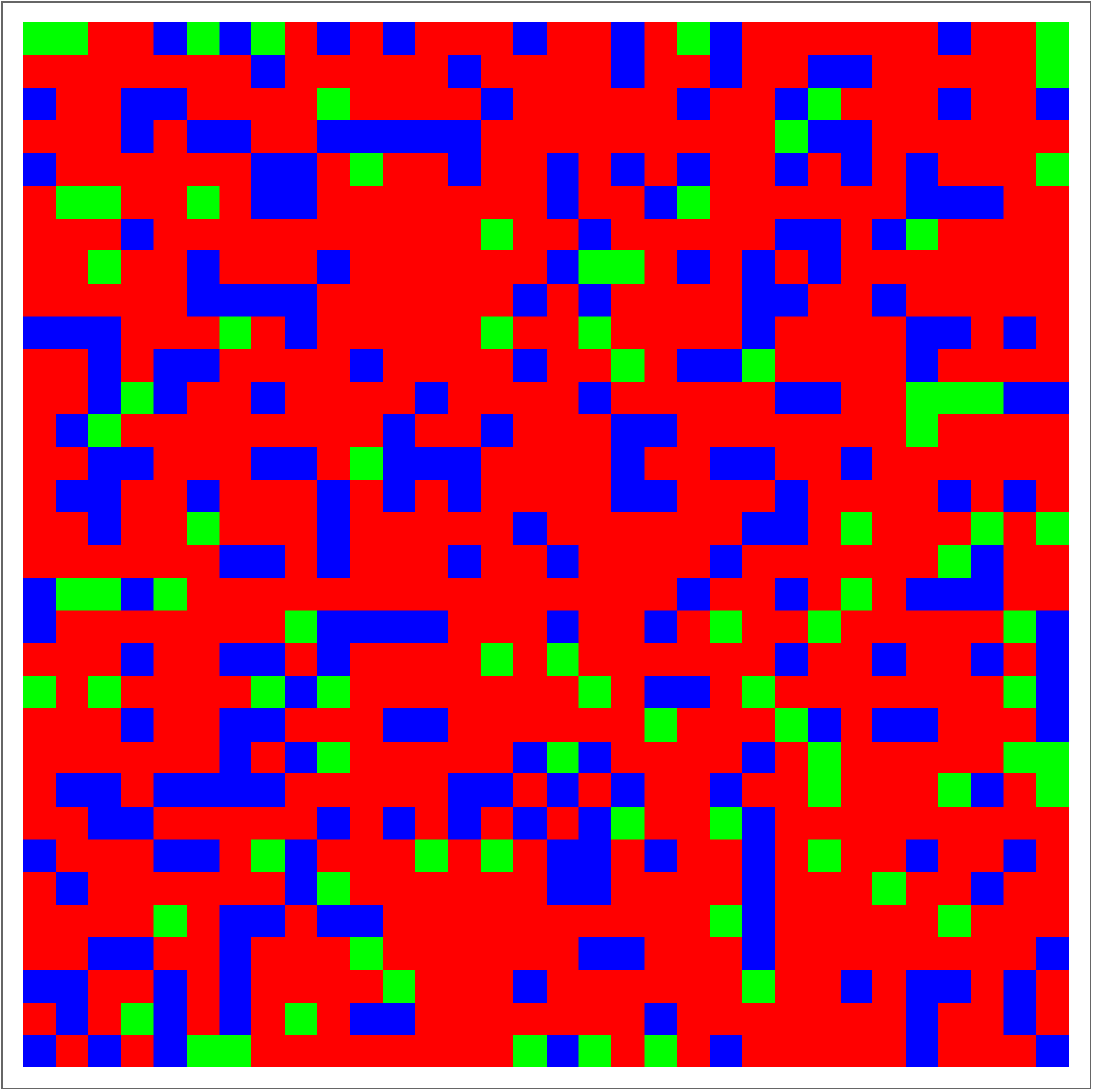}\quad
\includegraphics[width=.4\textwidth]{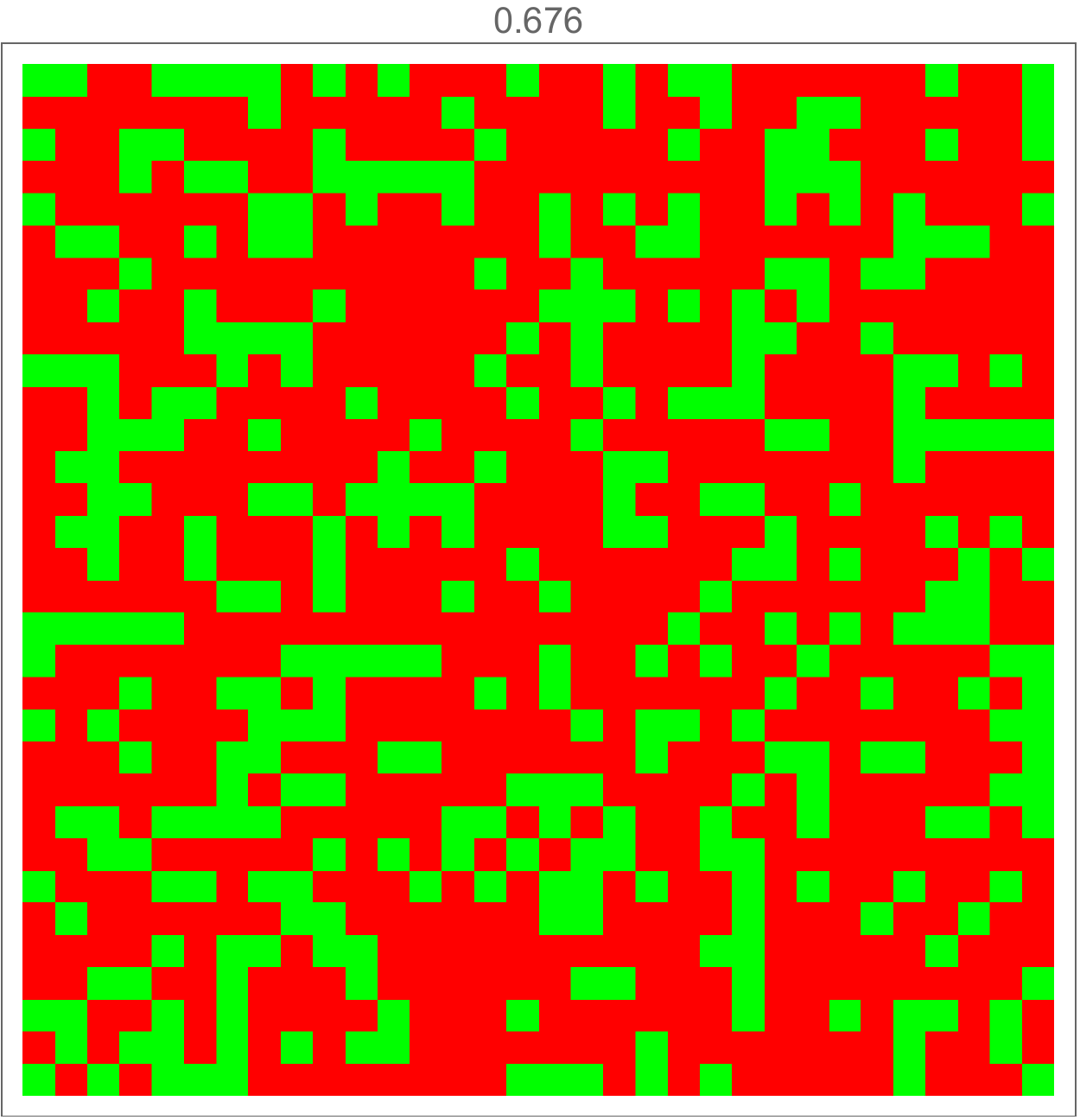}\quad
\includegraphics[width=.4\textwidth]{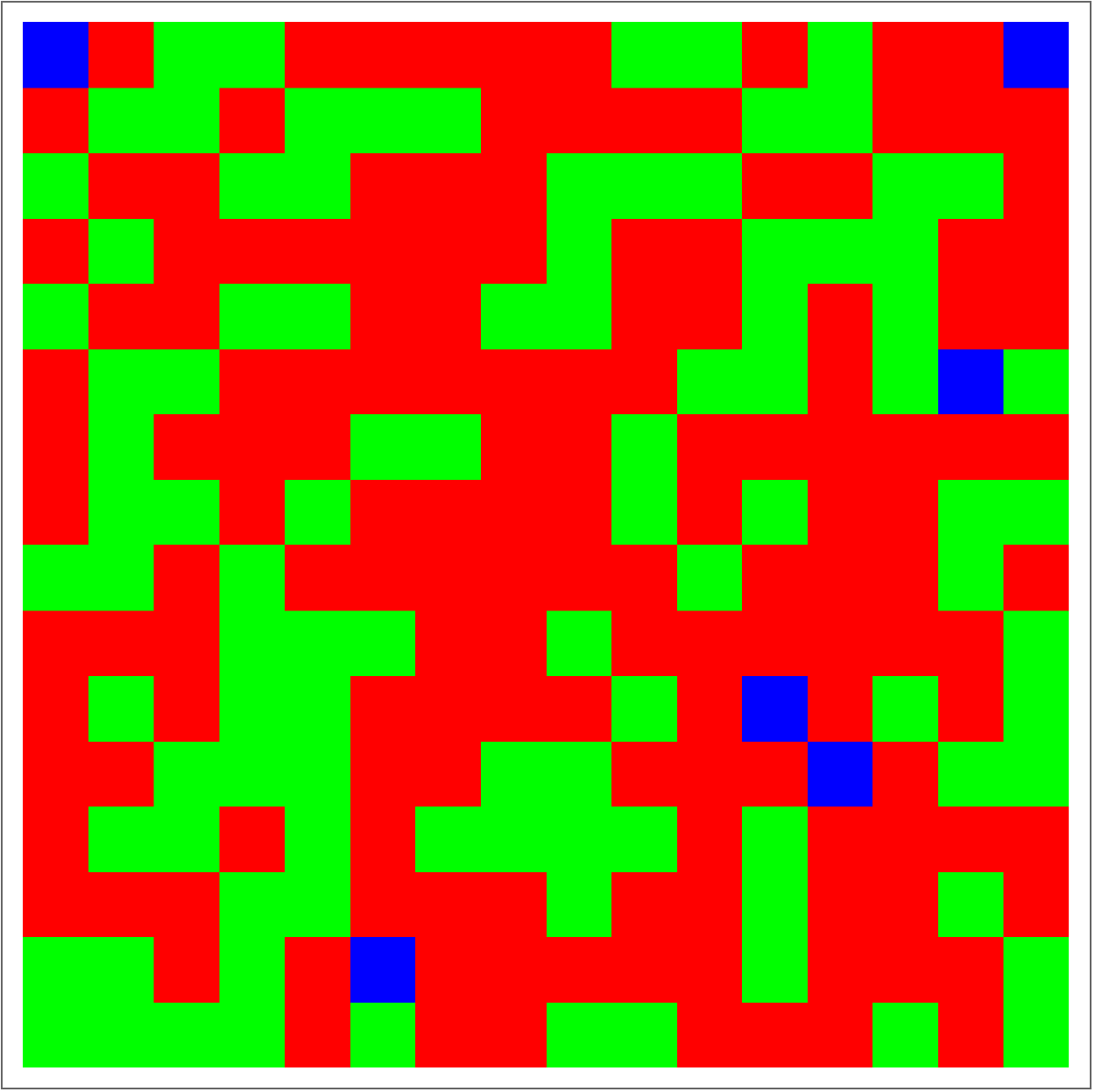}\quad
\includegraphics[width=.4\textwidth]{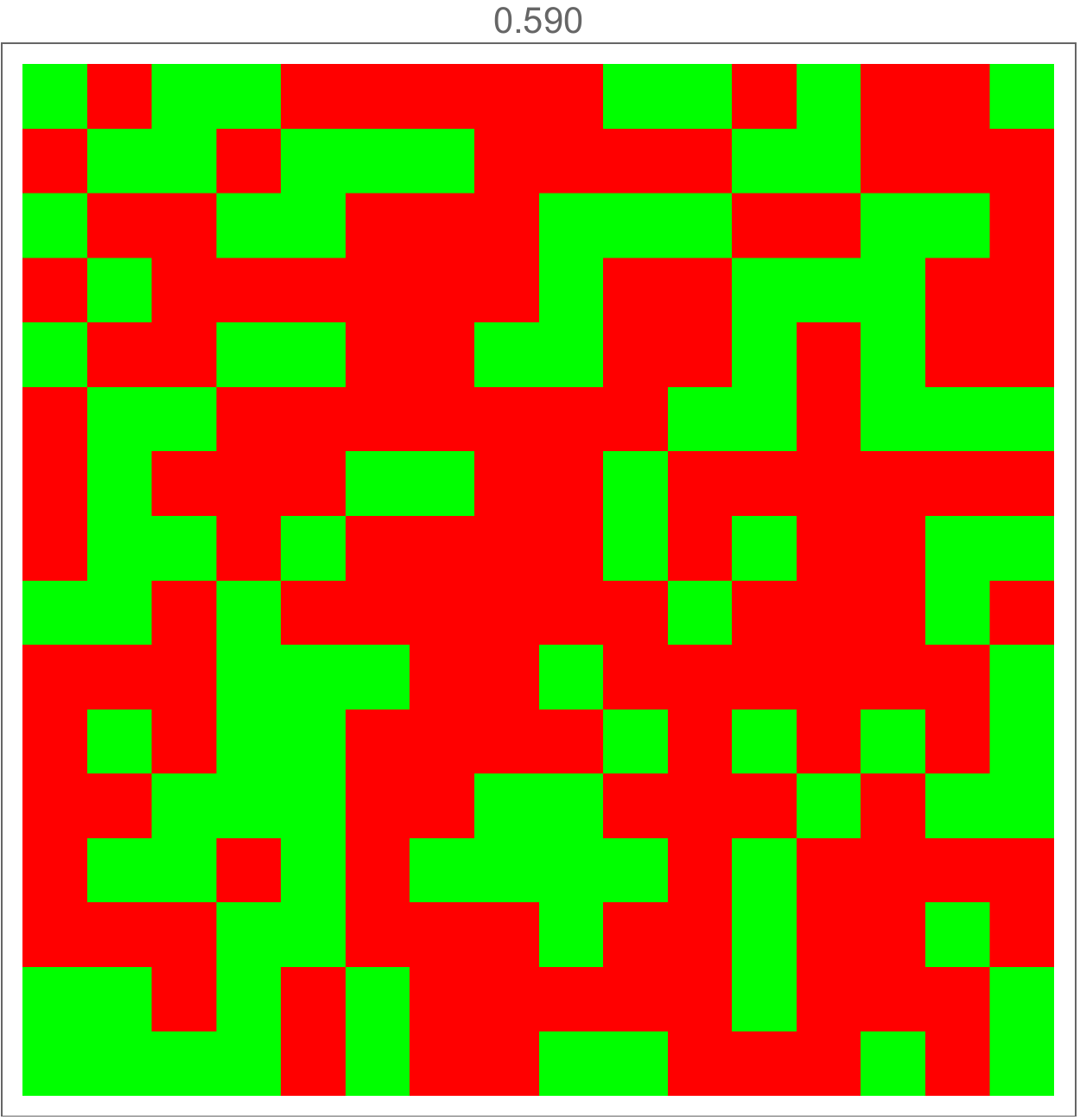}\quad
\includegraphics[width=.4\textwidth]{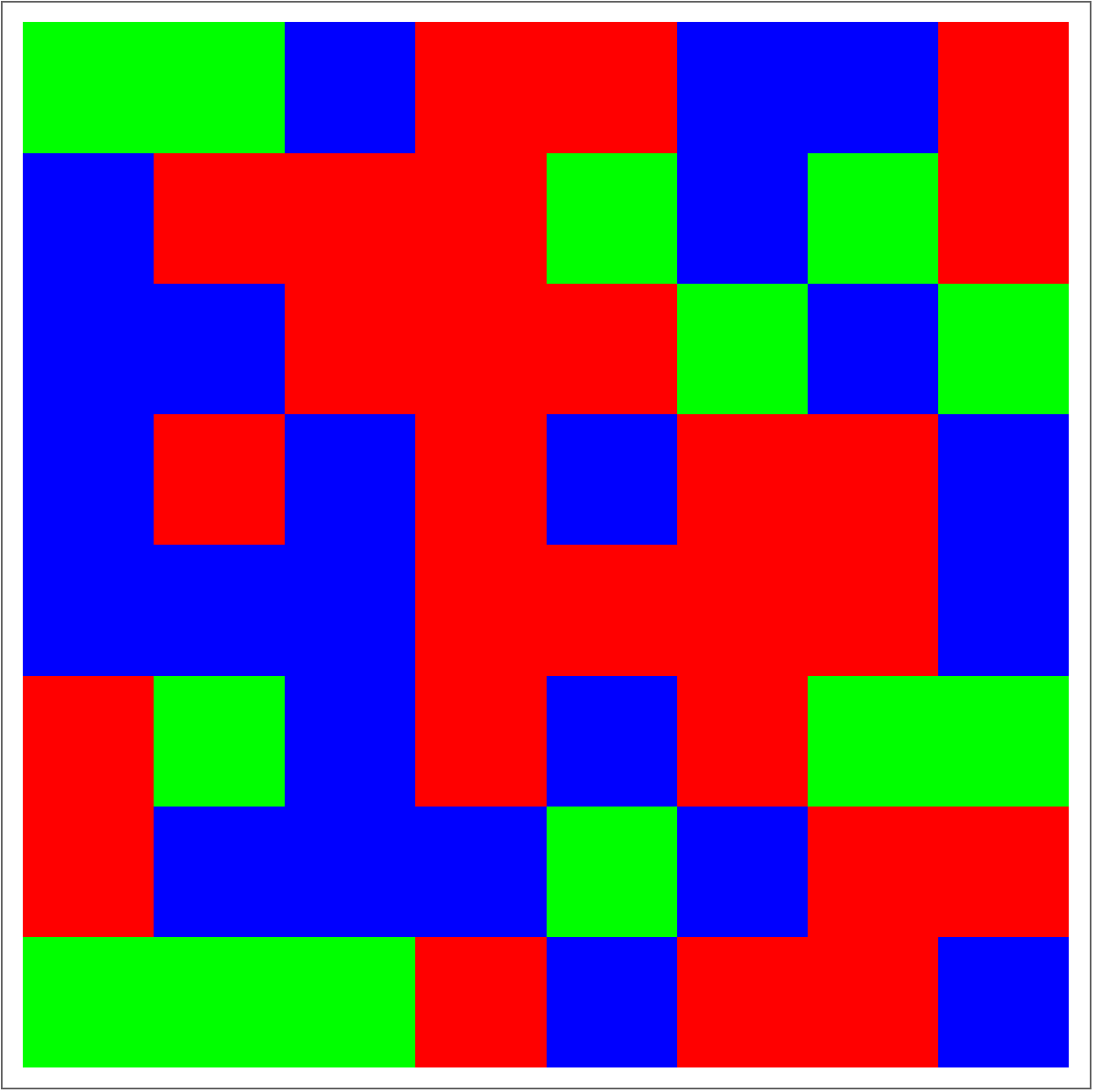}\quad
\includegraphics[width=.4\textwidth]{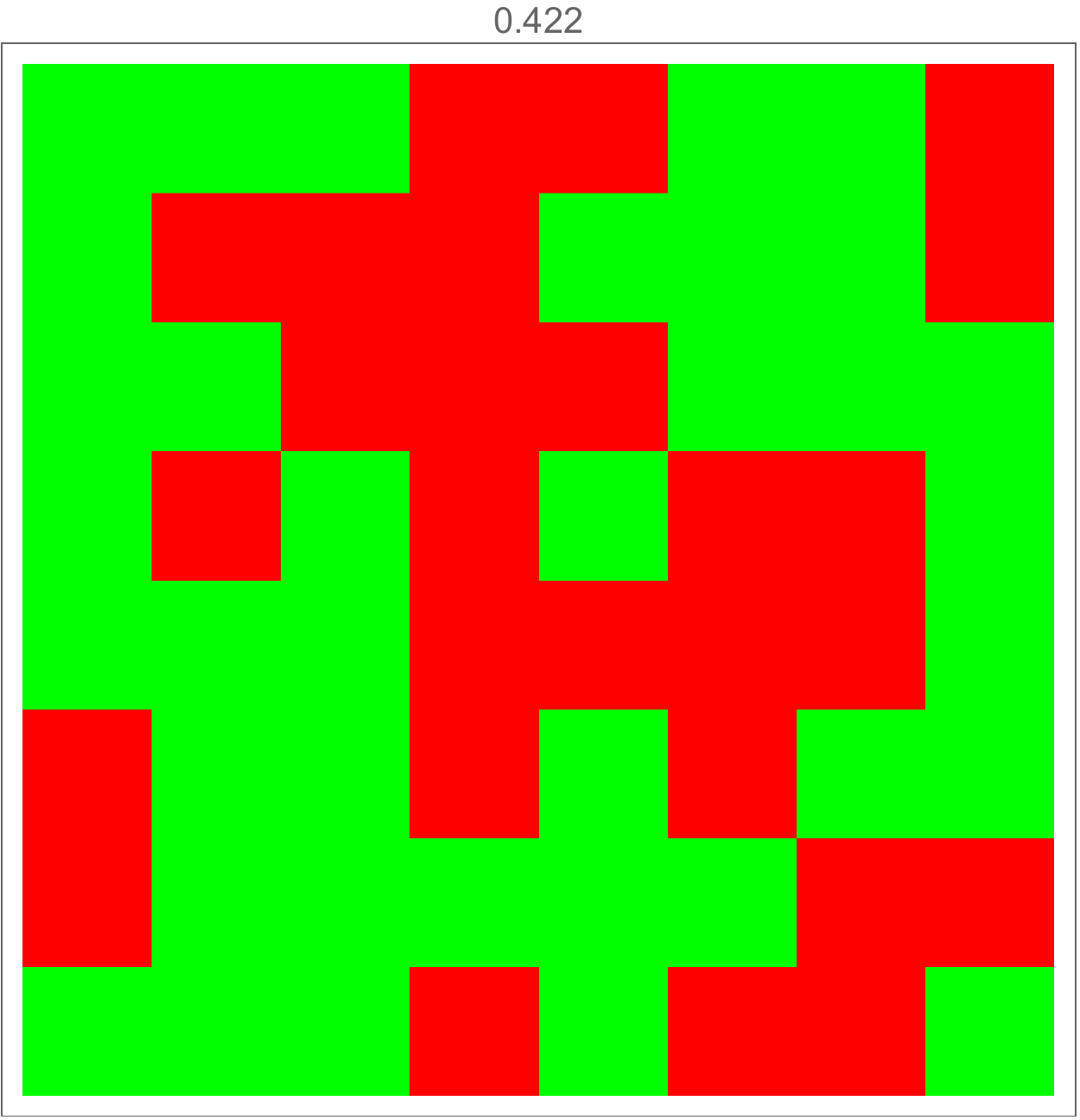}\quad
\caption{Continuation of Figure (\ref{f012}): third iteration with $p_3 = 0.676$ (top), fourth iteration with $p_4 = 0.590$ (middle), fifth iteration with $p_5 = 0.422$ (bottom).}
\label{f345} 
\end{figure} 

\begin{figure}
\centering
\includegraphics[width=.4\textwidth]{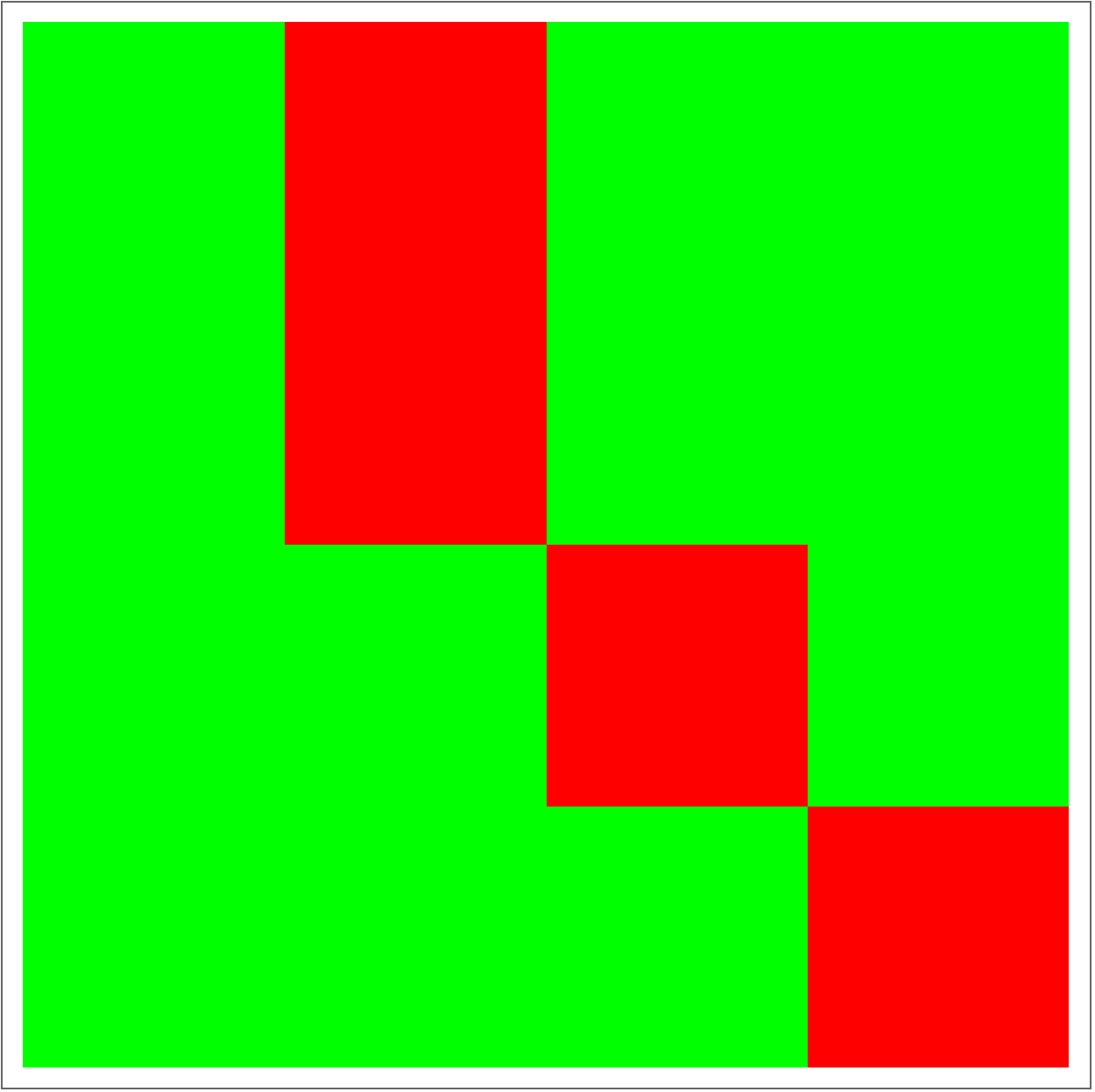}\quad
\includegraphics[width=.4\textwidth]{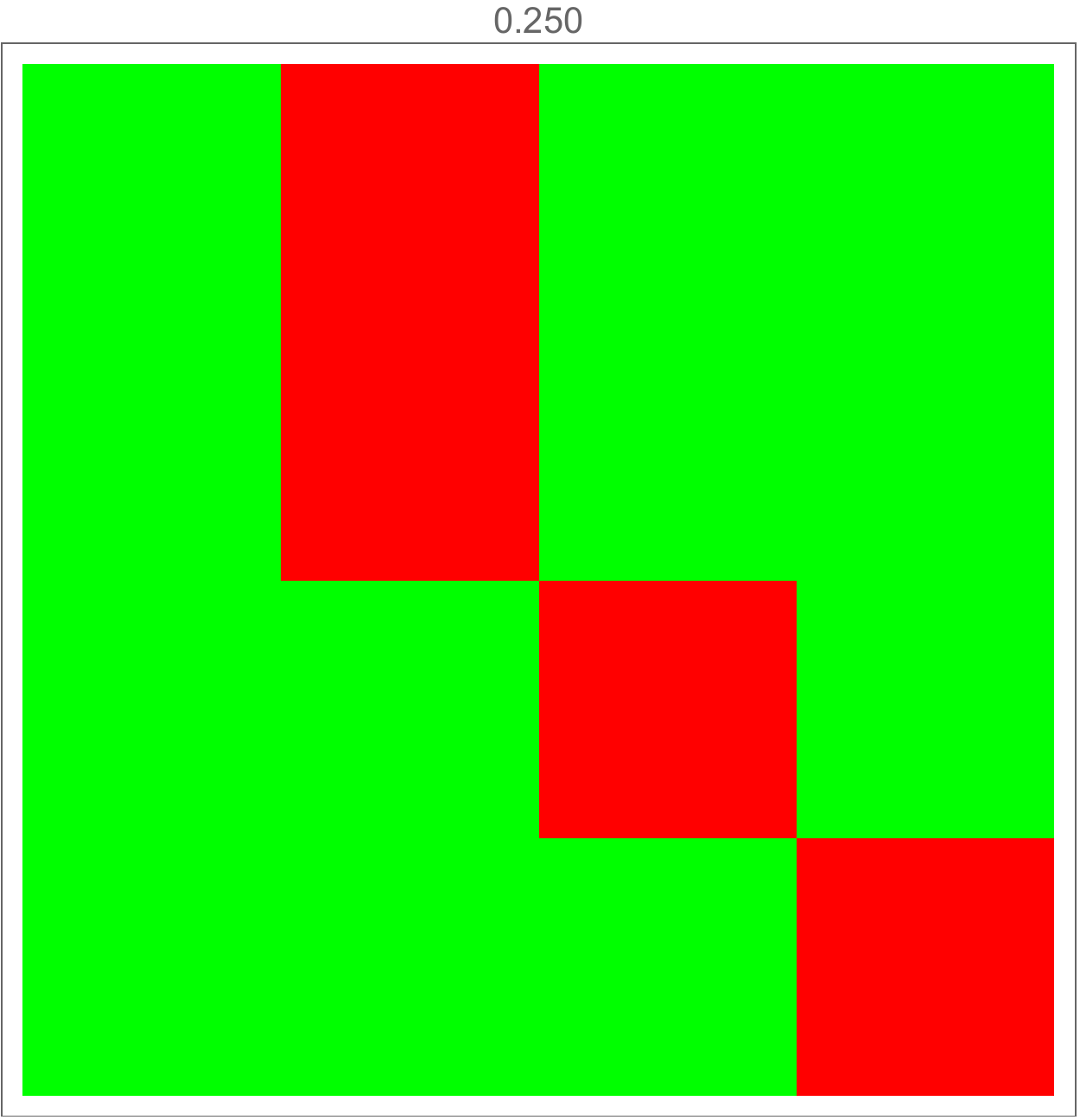}\quad
\includegraphics[width=.4\textwidth]{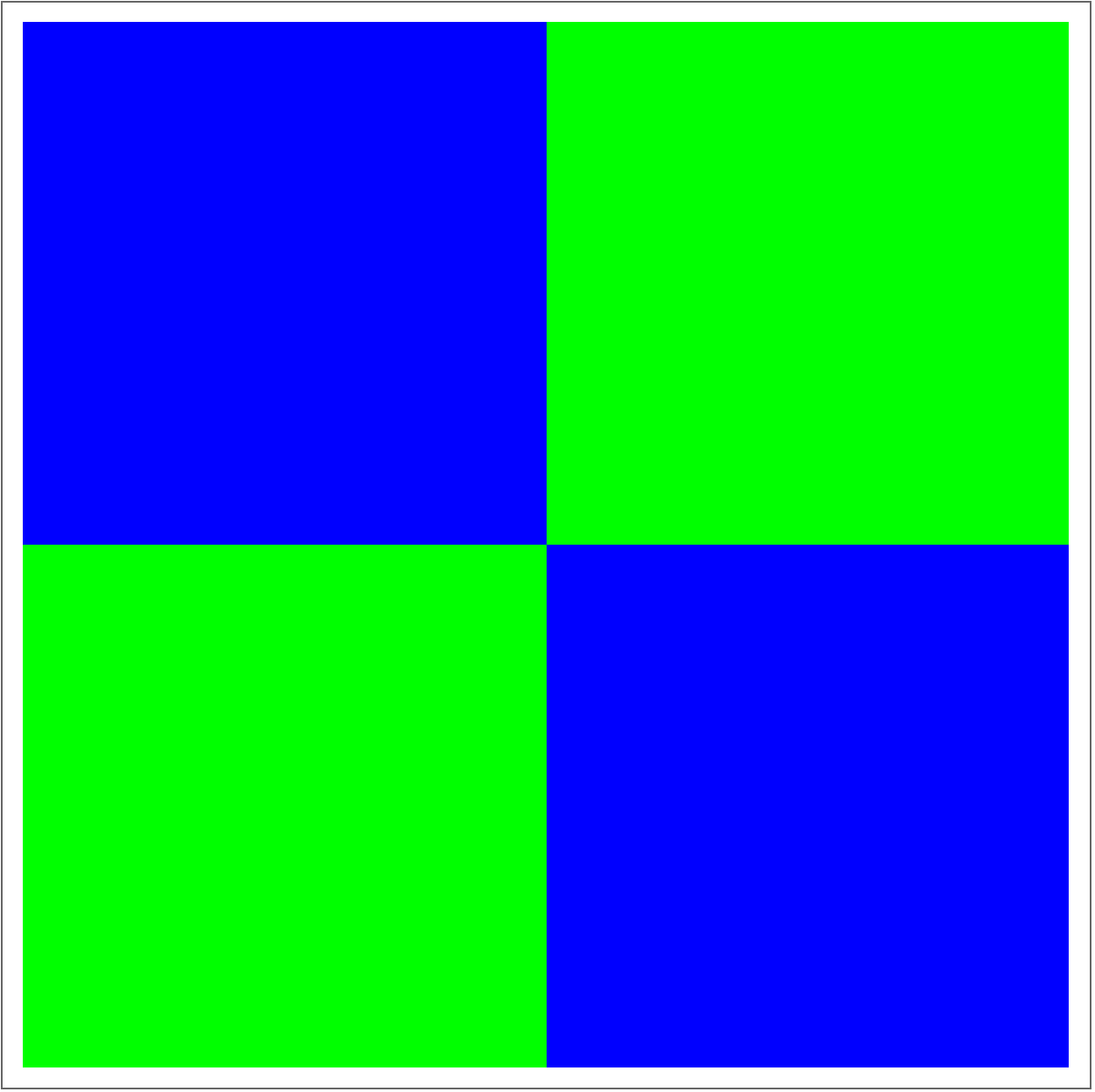}\quad
\includegraphics[width=.4\textwidth]{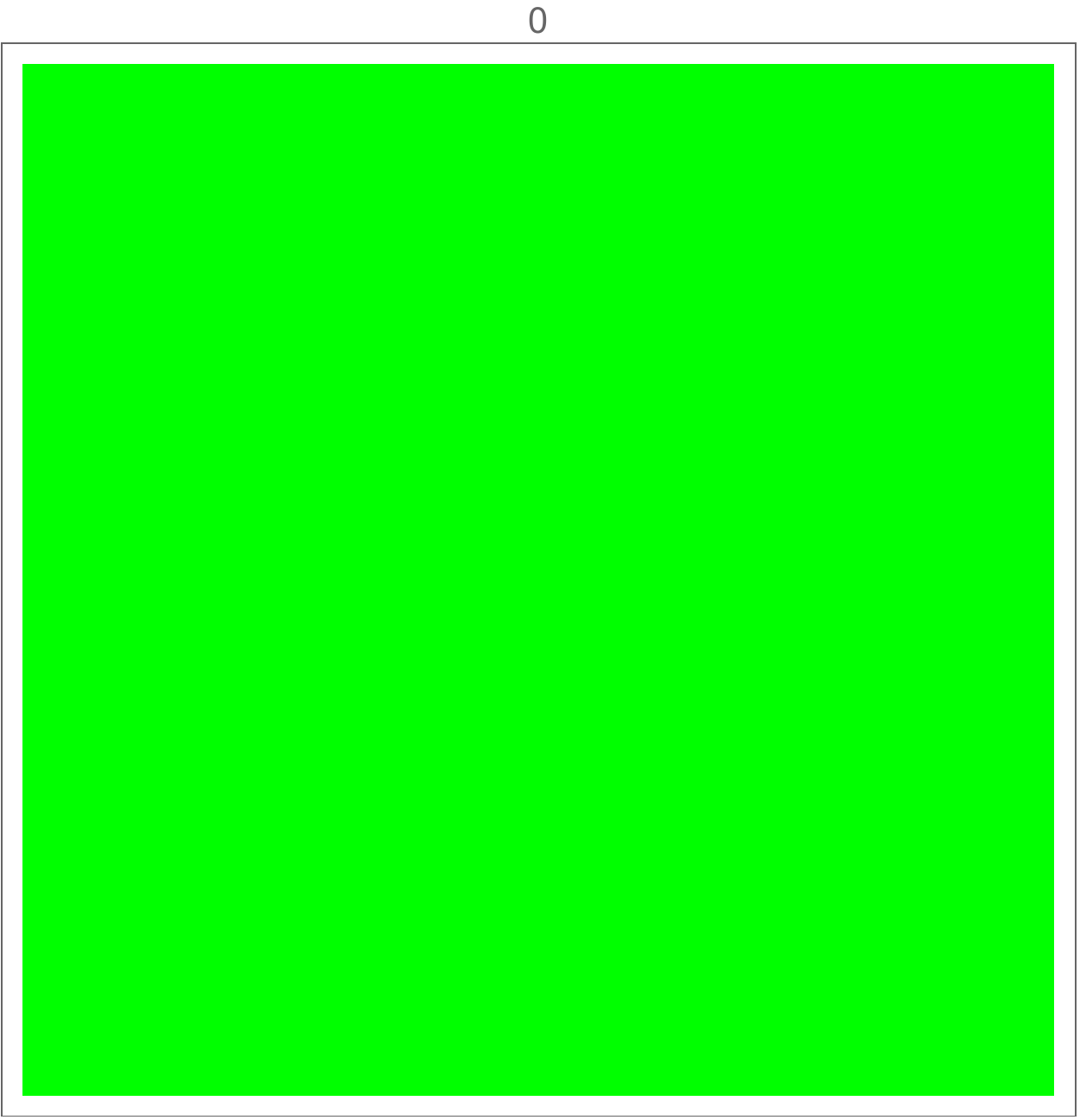}\quad
\includegraphics[width=.4\textwidth]{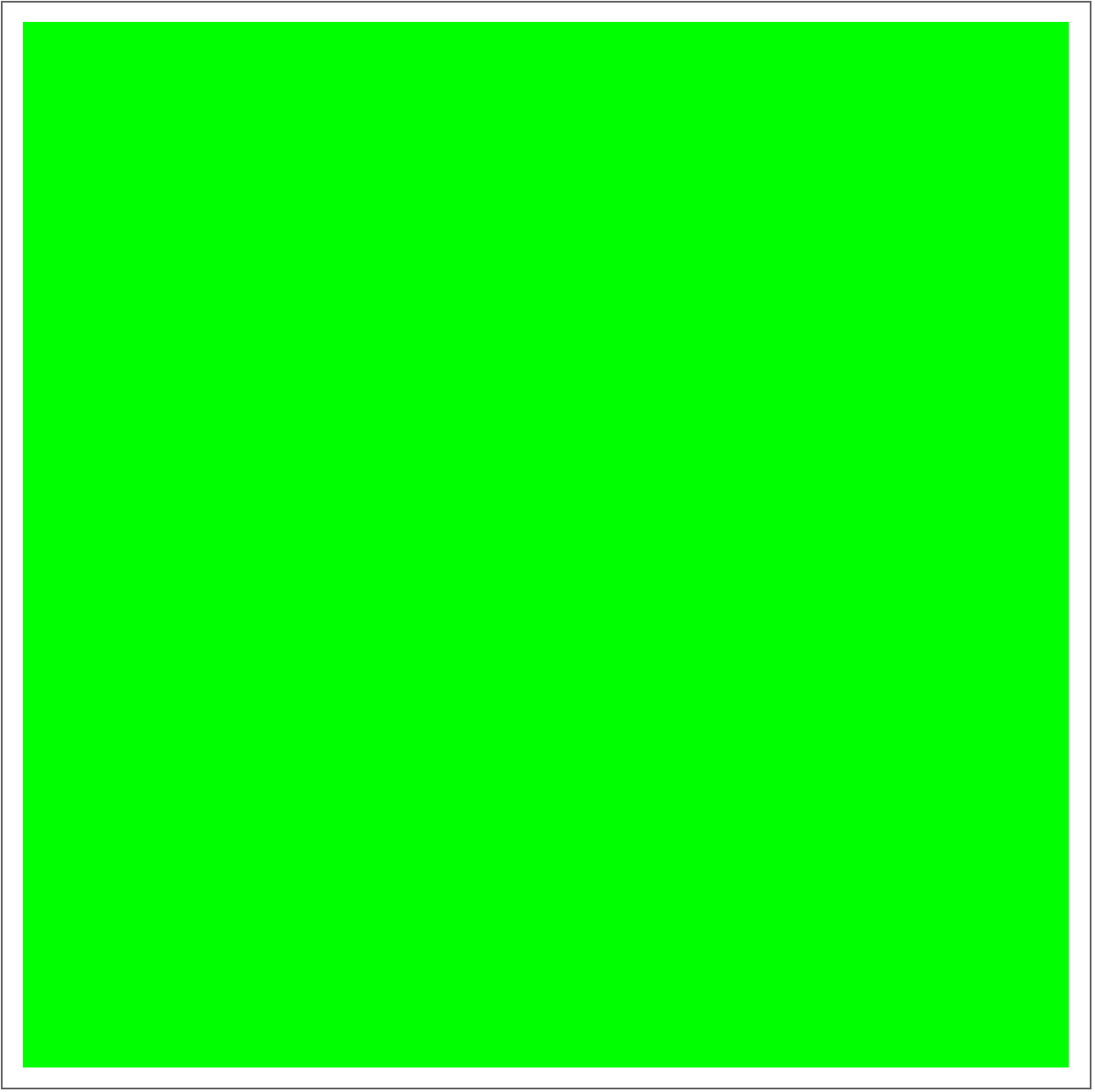}\quad
\includegraphics[width=.4\textwidth]{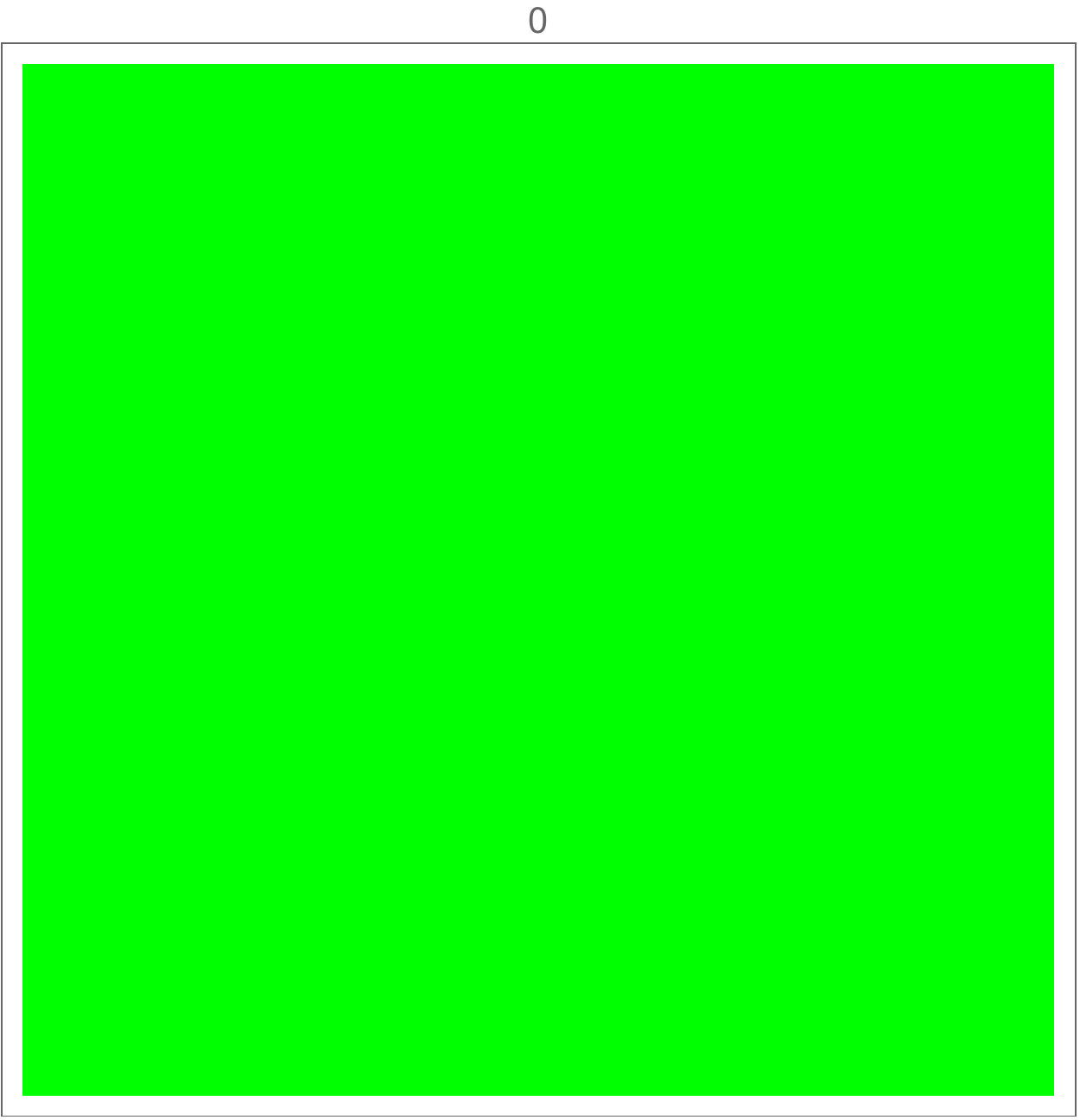}
\caption{Continuation of Figure (\ref{f345}): sixth iteration with $p_6 = 0.250$ (top), seventh iteration with $p_7 = 0.00$ (middle), eighth iteration with $p_0 = 0.000$ (bottom).}
\label{f678}
\end{figure} 

The series of Figure (\ref{f012},\ref{f345},\ref{f678}) shows how the various steps of the processing, which ends with a giant item labelled TF, which means the related person is wrongly tagged as NWT while being WT.

On the other hand, applying the presumption of guilt to uncertain aggregates would have lead to the right label within only 2 iterations with $0.750 \rightarrow 0.949 \rightarrow 1$.

In above case, 7 iterations reach a deterministic label TF, i.e. before applying the eight iteration, which encompasses all the 65536 ground items. In comparison, Eq. (\ref{m}) yields $m=9$ and iterations of Eq. (\ref{p1}) with $k=0$ gives the series 0.750, 0.738, 0.718, 0.684, 0.623, 0.516, 0.336, 0.114, 0.005, 0.000 in agreement with $m=9$. 

These discrepancies are consistent with fluctuations in the initial random distribution of the 65536 ground items. For instance,  $p_7=0.114$ indicates the possibility to reach the right label in $0.114\%$ of initial random distributions of the 65536 ground items with still  $p_0=0.750$. However, here the right label means the wrong mathematical result of the procedure since $p_0=0.750<p_{c,0}\approx 0.77$ implies to reach the attractor $p_{NWT}=0$. 

With eight iterations $p_8=0.005$, i.e. only five out of 1000 random distributions of the 65536 ground items with $p_0=0.750$ yields the mathematical wrong label TC. To ensure to get the mathematical right TF label requires to implement $m=9$ iterations, which here would imply to collect $4^8-4^9=196610$ ground items.


\section{The cases of Libya and Iraq}

Assuming that the processing of a dataset collected by monitoring an individual is similar to that used when monitoring a country about  some military issues, past cases of Libya and Iraq can be revisited in light of above findings. For both countries, the aim was to check the existence or not of Weapons of Mass Destruction (WMD). 

It happened that for Libya several National Intelligence agencies had wrongly agreed that there were no WMD \cite{lyb1,lyb2}. The model would point that all related countries were a priori convinced that no WMD were present and thus applied the presumption of innocence at uncertain aggregates of data.

At odd, the case of Iraq has been more singular with different National Intelligences reaching opposite conclusions. France and Russia had concluded that there were no WMD while America and England concluded the opposite \cite{ira}. 

The Americans and the British were convinced Iraq has WMD and were looking for proofs to validate their conviction. In contrast, the French and the Russians were convinced of the opposite. Accordingly, the American and the British would tag TC uncertain aggregates of data, while the French and the Russians would tag TF these uncertain aggregates. Having at hand similar data, my model would thus lead to opposite conclusions. 

This hypothesis does not imply an a priori bias from the various countries but evokes simply the way local uncertainties have been processed favoring a presumption of guild or innocence.

Therefore, reaching opposite conclusions using similar ground data sets could have been the "natural" result of the opposite expectations of the different protagonists without the need to distort the data. I do not claim that this is what happened, but only to show that an alternative explanation is plausible besides the predominant one of malicious data manipulation \cite{lie}.

\section{Conclusion}

In this work I have addressed the issue of processing data collected by monitoring a person to check a possible terrorist activity. However, the issue being both highly sensitive and confidential, I studied a mirror model, which in turn could shed some new light on the real issue.
The model considers a collection of data, the ground items, obtained by monitoring a given person with these data being labeled either Terrorist Connected (TC) or Terrorist Free (TF).

My goal was to investigate a possible invisible flaw embedded in the data processing itself, which would systematically distort the results.
And indeed, my analysis unveiled the existence of incompressible errors in diagnosing a would-be terrorist (WT) due to embedded biases in the processing of large amount of data. The errors occur systematically but only within a specific range of the proportion of TC ground items.

The resulting flaw proves to be irremovable because it is anchored in the very procedure of data processing in connection with the inevitable local appearances of indeterminate aggregates during its implementation. These uncertain aggregates are then naturally labeled as terrorist free (TF) in tune with the presumption of innocence. They are shown by the equations to drive the wrong systematic labelling always at the benefit of a WT, who are labeled not a would-be terrorist (NWT). Out of this range, the processing yields the right label.

On the other hand, applying the presumption of guilt TC to the uncertain aggregates is found to shift the systematic errors to another area of proportions of TC items. There, no WT is wrongly labelled NWT. However, NWT are wrongly labeled WT.

My mirror model is very simple and it may well be disconnected from the reality I claim to address, but yet it unveils a track for systematic failures, which in any case could be worth to investigate by intelligence agencies. The focus of an investigation should be the overall effect of "natural" and  inconsequential biases applied locally during the data processing.

In case the results have some relevance to reality, citizens should be made aware of the ethical choice of presumption of innocence that is applied in case of uncertain data, in conformity with the values of the rule of law in democratic countries. This would avoid public misunderstanding when other failures inevitably occur, such as during the dramatic terrorist attacks in Paris (2015) and Brussels (2016).

Finally, one of the referees mentioned that my hierarchical scenario for dealing with the data resembles the random forest algorithm, and I intend to explore this avenue in the near future  \cite{forest}. Indeed, it would be fruitful to interest computer scientists working on intelligence to further develop my model. Hopefully, future investigation may reveal paths to reduce the range of incompressible errors and even to their eventual eradication, which could be at play in the current procedures used to process large amount of heterogeneous data.


\end{document}